\documentclass[preprint2]{aastex}

\newcommand\Rsun{\hbox{$R_\odot$}}

\newcommand\Msun{\hbox{$M_\odot$}}
\newcommand\Rstar{\hbox{$R_*$}}

\newcommand\TL{\hbox{$T_L$}}
\newcommand\fir{\hbox{$fir$}}

\newcommand\etal{\hbox{et~al.}}
\newcommand\ROSAT{\hbox{\it ROSAT}}
\newcommand\ASCA{\hbox{\it ASCA}}
\newcommand\Chandra{\hbox{\it Chandra}}
\newcommand\Einstein{\hbox{\it Einstein}}

\newcommand\OVI{\hbox{O {\sc vi}}}
\newcommand\OVII{\hbox{O {\sc vii}}}
\newcommand\OVIII{\hbox{O {\sc viii}}}
\newcommand\MgXI{\hbox{Mg {\sc xi}}}
\newcommand\MgXII{\hbox{Mg {\sc xii}}}
\newcommand\FeXVII{\hbox{Fe {\sc xvii}}}
\newcommand\NeIX{\hbox{Ne {\sc ix}}}
\newcommand\NeX{\hbox{Ne {\sc x}}}
\newcommand\SiXIII{\hbox{Si {\sc xiii}}}
\newcommand\SiXIV{\hbox{Si {\sc xiv}}}
\newcommand\SXV{\hbox{S {\sc xv}}}
\newcommand\SXVI{\hbox{S {\sc xvi}}}
\newcommand\ArXVII{\hbox{Ar {\sc xvii}}}
\newcommand\CaXIX{\hbox{Ca {\sc xix}}}

\newcommand\FeXX{\hbox{Fe {\sc xx}}}
\newcommand\FeXXII{\hbox{Fe {\sc xxii}}}
\newcommand\FeXXIII{\hbox{Fe {\sc xxiii}}}
\newcommand\FeXXIV{\hbox{Fe {\sc xxiv}}}
\newcommand\FeXXV{\hbox{Fe {\sc xxv}}}
\newcommand\zori{\hbox{$\zeta\ ${\rm Ori}}}
\newcommand\zpup{\hbox{$\zeta\ ${\rm Pup}}}
\newcommand\COB{\hbox{Cyg OB2}}

\newcommand\cygate{\hbox{Cyg OB2 No. 8a}}

\newcommand\dori{\hbox{$\delta\ ${\rm Ori}}}
\newcommand\tsco{\hbox{$\tau\ ${\rm Sco}}}
\newcommand\toriC{\hbox{$\theta^1$ Ori C}}
\newcommand\Berghofer{\hbox{Bergh\"{o}fer}}
\newcommand\vinf{\hbox{$v_\infty$}}
\newcommand\Mdot{{\hbox{$\dot M$}}}
\newcommand\Msunyr{\hbox{$M_\odot\,$yr$^{-1}$}}
\newcommand\kms{\hbox{km$\,$s$^{-1}$}}
\newcommand\LxLbol{{\hbox{$L_x/L_{bol}$}}}

\newcommand\lamz{\hbox{$\lambda_{\circ}$}}
\newcommand\linflux{\hbox{line flux$/10^{-13}$}}

\newcommand{\vsini}{$v \sin i$}
\newcommand\Rfir{\hbox{$R_{fir}$}}

\makeatletter
\renewcommand{\section}{\@startsection%
{section}{1}{0mm}{-\baselineskip}%
{0.5\baselineskip}{\normalfont\Large\bfseries}}%
\makeatother

\shorttitle{X-ray lines from Cyg OB2 stars}

\begin{document}

\title{High Resolution X-ray Spectra of the Brightest OB Stars in the 
Cygnus OB2 Association}

\author{W. L. Waldron\altaffilmark{1}, J. P. Cassinelli\altaffilmark{2},
N. A. Miller\altaffilmark{3}, J. J. MacFarlane\altaffilmark{4}, and J. C.
Reiter \altaffilmark{3}}
\altaffiltext{1}{L-3 Communications Government Services, Inc.,
1801 McCormick Dr., Suite 170, Largo MD 20774; wayne.waldron@L-3com.com}
\altaffiltext{2}{Astronomy Department, University of Wisconsin-Madison, 475
N. Charter St., Madison, WI  53706; cassinelli@astro.wisc.edu}
\altaffiltext{3}{Physics and Astronomy Department, University of
Wisconsin-Eau Claire, PO Box 4004 Eau Claire WI 54702; millerna@uwec.edu;
reiterjc@uwec.edu} \altaffiltext{4}{Prism Computational Sciences, 16 N.
Carroll St. Madison WI 53703; jjm@prism-cs.com}

\begin{abstract}

The Cygnus OB2 Association contains some of the most luminous OB stars in
our Galaxy and the brightest of which are also among the most luminous in
X-rays.  We obtained a \Chandra\ High Energy Transmission Grating
Spectrometer (HETGS) observation centered on Cyg OB2 No. 8a, the most
luminous X-ray source in the Association.  Although our analysis will focus
on the X-ray properties of Cyg OB2 No. 8a, we also present limited analyses
of three other OB stars (Cyg OB2 Nos. 5, 9, and 12).  Applying standard
diagnostic techniques as used in previous studies of early-type stars (e.g.,
Waldron \& Cassinelli 2001), we find that the X-ray properties of Cyg OB2
No. 8a are very similar to those of other OB stars that have been observed using high-resolution
X-ray spectroscopy. From analyses of the He-like ion \fir\ emission lines
(\MgXI, \SiXIII, \SXV, and \ArXVII), we derive radial distances of the He-like line emission
sources and find that the higher energy ions have their lines form closer to the
stellar surface than those of lower ion states. Also these
\fir-inferred radii are found to be consistent with their corresponding X-ray
continuum optical depth unity radii.  Both of these findings are in
agreement with previous O-star studies, and again suggests that anomalously
strong shocks or high temperature zones may be present near the base of the
wind. The observed X-ray emission line widths ($HWHM \sim 1000$ \kms) are
also compatible with the observations of other O-star supergiants.  Since
Cyg OB2 No. 8a is similar in spectral type to \zpup\ (the only O-star which
clearly shows asymmetric X-ray emission line profiles with large
blue-shifts), we expected to see similar emission line characteristics.
Contrary to other O-star results, the emission lines of Cyg OB2 No. 8a show a large range in line
centroid shifts ($\sim$ -800 to +250 \kms).  However, we argue that most of the largest shifts
may be unreliable, and the resultant range in shifts is much less than those observed in \zpup. 
Although there is one exception, the H-like \MgXII\ line which shows a blue-shift of -550 \kms,  
there are problems associated with trying to understand the nature of this isolated large
blue-shifted line.  To address the degree of asymmetry in these line
profiles, we present Gaussian best-fit line profile model spectra from \zpup\ to illustrate the
expected asymmetry signature in the $\chi^{2}$ residuals.  Comparisons of the Cyg OB2 No. 8a
best-fit line profile residuals with those of \zpup\ suggest that there are no indications of any
statistical significant asymmetries in these line profiles.  Both the line shift characteristics and lack
of line asymmetries are very puzzling results.  Given the very high mass loss rate of Cyg OB2 No.
8a (approximately five times larger than previous \Chandra\ observed O supergiants), the emission
lines from this star should display a significant level of line asymmetry and blue-shifts as compared
to other OB stars.  We also discuss the implications of our results in light of the fact that Cyg
OB2 No. 8a is a member of a rather tight stellar cluster, and shocks could arise at interfaces with
the winds of these other stars.
\end{abstract}


\pagestyle{plain}
\pagenumbering{arabic}

\section{Introduction}

The Cygnus OB2 association (\COB) is significant in the history of stellar
X-ray astronomy in that \Einstein\ Observatory observations of \COB\ were
the first to show that early-type stars are X-ray sources (Harnden \etal\
1979). Cassinelli \&\ Olson (1979) predicted that OB stars should be X-ray
sources based on the superionization stages seen in UV spectra, and
suggested that the X-rays arose from a coronal region at the base of a cool
wind. Based on their Cyg OB2 X-ray observations, Harnden \etal\ found that
the coronal plus cool wind model predicted too low a flux of soft X-rays
owing to the expected attenuation by the overlying wind, and they suggested
that the X-rays come from a distribution of shocks in the stellar
wind. Lucy \&\ White (1980) developed the first shock model for these stars,
and the concept of embedded shocks has remained the prevailing picture
for the origin of OB stellar X-rays. The X-ray observations of Cyg
OB2 using \Einstein\ (Harden \etal\ 1979), \ROSAT\ (Waldron \etal\ 1998), and 
\ASCA\ (Kitamoto \&\ Mukai 1996) found that the brightest stars in this
association have significantly larger X-ray luminosities in comparison with
other OB stars. However, the \COB\ stars have larger bolometric
luminosities, and the X-ray to bolometric luminosity ratio is in line with
that of other OB stars (i.e., with an observed \LxLbol\ $\sim 10^{-7}$;
\Berghofer\ \etal\ 1997). Among the early type stars in Cyg OB2, star No. 8a
is particularly interesting because it is one of the most X-ray luminous
O-stars ($L_x=1.5 \times 10^{33}$ ergs~ s$^{-1}$), and because its fast wind
has a large mass loss rate, (\Mdot $ > 10^{-5}$ \Msunyr) which is almost as
large as that of Wolf-Rayet stars. In addition, all of the four brightest stars in \COB\ ( Nos. 5, 8a,
9, \& 12) are also know to be variable, non-thermal radio sources (Abbott,
Bieging, \& Churchwell 1981; Bieging, Abbott, \& Churchwell 1989).

Since this association contains some of the most luminous stars in our
galaxy (\COB\ Nos. 9 and 12), it has been the focus of many massive star
studies. For example, Kn\"{o}dlseder (2000), using the 2MASS near infrared
survey data, analyzed the extent and population properties of \COB\ and
proposed that it should be reclassified as a ``young globular cluster''
instead of an OB Association. As he points out, this association is
providing opportunities to study the upper end of the main sequence of stars
in a rather dense population.  The subject of massive star clusters has
recently become even more important owing to the well received idea that
massive stars are formed by mergers in tight clusters (Bonnell \& Bate 2002; Bonnell, Vine, \&
Bate 2004).  The core of Cyg OB2 might originally have been such a region.

The rapidly expanding stellar winds from OB stars are often considered one
of the most unexpected and important discoveries of NASA's early space
program (Snow \& Morton 1977). Over the years we have learned a great deal
about the wind driving mechanisms and wind properties (see Lamers \&\
Cassinelli 1999). However, it is the nature of the processes that lead to
shocks and X-ray emission that is still unclear and is currently the center
of attention in OB stellar wind research. As shown by recent studies
(Waldron \&\ Cassinelli 2001; Cassinelli \etal\ 2001; Miller \etal\ 2002),
the resolving power of the \Chandra\ High Energy Transmission Grating
Spectrometer (HETGS) can provide the critical information required to help
us understand these processes.  For the first time, astronomers are observing
spectrally resolved X-ray line profiles from OB stars. In particular,
tight line complexes like the forbidden, intercombination, and
resonance (\fir) lines of He-like ions are spectrally resolved by the HETGS.
There are relatively few OB stars that will probably be observed with
\Chandra\ at this high spectral resolution since large exposure times are
required, so it is important to glean as much information as possible from
targets such as the \COB\ stars.

Before the launch of \Chandra\ there was only indirect evidence for a
shocked-wind origin for OB stellar X-rays. These include: a) the lack of
sufficient oxygen K-shell absorption in previous lower resolution X-ray
spectra (Cassinelli \etal\ 1981; Cassinelli \& Swank 1983; Corcoran \etal\
1993); b) the consistency of observed \OVI\ UV P-Cygni line profiles with an
embedded X-ray source (MacFarlane \etal\ 1993); and c) theoretical wind
instability studies (Lucy \& White 1980; Owocki, Castor, \& Rybicki 1988; Cooper 1994;
Feldmeier 1995). In our \Chandra\ observations of \zori\ (Waldron \&\
Cassinelli 2001), \zpup\ (Cassinelli \etal\ 2001), and \dori\ (Miller \etal\
2002), we found that far more precise information regarding the {\it
location} of the X-ray sources could be determined from analyses of the
He-like ion \fir\ lines (e.g., \OVII, \NeIX, \MgXI, \SiXIII, \SXV).  In
particular, the $f/i$ line ratio has long been used as an electron density
diagnostic for solar-like plasmas. However, for the case of OB stars, it is
the large mean intensity of UV/EUV photons, and not the density of
electrons, that determines the relevant population levels, as discussed by
Blumenthal, Drake, \& Tucker (1972).  The $f/i$ line ratio can be used to
derive the geometrical dilution factor of the stellar radiation field, and
this allows us to extract the distance between the \fir\ line formation
region and the photospheric source of UV/EUV radiation. We call \Rfir\ the
\fir-inferred radius of the line formation region derived from the He-like
ions.

Our analyses indicate that the derived \Rfir\ for the O-stars correspond
reasonably well with their respective X-ray continuum optical depth unity
wind radii as summarized by Waldron \& Cassinelli (2002).  This
correspondence in radii indicates that the X-rays source regions are
distributed throughout the wind from just above the photosphere out to
approximately 10 \Rstar. However, this does not mean that the entire wind is
emitting X-rays, but rather that there are sources of hot X-ray emitting
material embedded in much cooler gas at all levels of the wind.  In fact,
the X-rays must be arising from only a small fraction of the matter in the
wind.  For example, by assuming that the total {\it observed} X-ray emission
is approximately equal to the total {\it intrinsic} X-ray emission, the
resultant emission measures are found to be 4 to 5 orders of magnitude
smaller than the total wind emission measure (Cassinelli \etal\ 1981; Kahn
\etal\ 2001). This small fraction of very hot gas is probably in the form of
shock fragments and filaments in the clumpy and turbulent outflows.

Although the results seem to support the idea that O-star X-rays arise in
shock fragments embedded in stellar winds, two interesting problems have
emerged.  First, the lines of the highest ion stages arise deep in the flow
and require shock jumps, $\Delta v$, that are larger than the local wind
speed (Waldron \& Cassinelli 2001).  Several suggestions have been offered:
fast ejecta emerge from sub-photosphere regions (Feldmeier, Shlosman, \&
Hamann 2002); in-fall of stalled wind material in the form of clumps (Howk
\etal\ 2000); or confinement of anomalously hot gas in magnetic loops
(Cassinelli \&\ Swank 1983; Waldron \&\ Cassinelli 2001; Schulz \etal\
2003). Second, the observed X-ray line profile shapes do not conform to pre-launch expectations.
As expected from an expanding distribution of X-ray sources embedded in a
stellar wind, the observed X-ray line profiles from O-stars with massive
winds show a large range in broadness with $HWHM$ of $\sim 200$ to $1500$ \kms.  However,
with the exception of one star, the line centroids are {\it
not} Doppler blue-shifted.  Such a shifting or skewing of line profiles to
shorter wavelengths was expected, as predicted by MacFarlane \etal\ (1991)
as an effect of X-ray absorption by the wind. The X-rays from the shocks on
the back-side (red-shifted) part of the wind should be more strongly
attenuated than the X-rays from the wind material on the front-side moving
towards the observer (blue-shifted). The O4f star \zpup\ (the only early
O-star spectral type observed so far) is currently the only star thus far
which clearly shows the expected blue-ward shifted skewed X-ray line
profiles, so it is of special interest to see if the O5.5 I(f) star \cygate\
shows similar line properties.

The lack of Doppler blue-shifted lines in nearly all OB stars is without a
doubt the most unexpected result obtained by \Chandra\ observations.
Although the simplest explanation for this dilemma is to say that all the
X-ray emission is located at radii where the wind attenuation of the
receding X-ray emission is negligible, the problem is that these radii are
found to be much larger (order of magnitude) than the \fir -inferred radii
(Waldron \&\ Cassinelli 2001).  Even for the largest observed \Rfir, the
wind density is still large enough such that most of the red-shifted line
emission should be attenuated, and blue-shifted, asymmetric line profiles
should still be observed. A possible explanation is that the winds are
especially porous or `clumped' which would allow significantly more
red-shifted X-rays from the back side to propagate through the stellar wind.
For example, the winds may have small filling factors of absorbing material
as would be the case for highly clumped winds.  The idea of highly clumped
winds in OB stars is gaining support. Waldron \&\ Cassinelli (2001)
presented the first attempt to fit HETGS line profiles with a wind
distributed X-ray source model and concluded that the wind absorption had to
be significantly reduced either by large reductions in the mass loss rate or
the X-rays are distributed in a highly non-symmetric wind (i.e., clumps).
Even for the only star showing clear blue-shifts and asymmetric lines (i.e.,
\zpup), Kramer, Cohen \& Owocki (2003) also found that significant
reductions in the wind absorption were required to explain these profiles.
Howk \etal\ (2000) present arguments that the X-rays from OB stars are
formed in bow shocks around clumps in the winds.  The fact that shocks lead
to highly compressed regions that may be Rayleigh-Taylor unstable led
Feldmeier, Oskinova, \& Hamann (2003) to consider the effects of absorption in fragmented
winds.  They show that the fragmented nature of a wind allows the X-rays to
escape from much deeper in the wind than would be the case for smooth,
un-clumped winds. Interestingly, the idea that clumps could be important for
the formation and transmission of X-rays in OB stellar winds can be traced
back to Lucy \& White (1980) and their first alternative model to the
coronal idea for X-ray formation. The Lucy \& White (1980) model had fast
clumps being driven by radiation forces through the ambient wind and this
would produce frontal (or forward facing) bow shocks.  The more recent
models such as Howk \etal\ (2000) have the fast winds colliding with slower
clumps and these produce reverse shocks. Reverse shocks also tend to be the
dominant source of X-rays in one-dimensional hydrodynamic models (e.g.,
Feldmeier 1995).

Although wind clumping appears to be a possible answer to the blue-shift
problem, the clumps introduce a dilemma with regards the $f/i$ line ratio
analysis results.  Since a highly clumped wind implies that the mass loss rates reported for these
stars are actually overestimates of their true values, then the relationship between the
\fir-inferred radii and X-ray continuum optical depth unity radii is no
longer correct, and we are then faced with a new problem of understanding
the significance of this observed correlation.  In addition, whatever level
of wind clumping is required to explain the X-rays must also be consistent
with the observational constraints established at other wavelength bands
(e.g., UV, IR, \&\ radio). Again, historically, it is interesting to note
that the main argument against the base coronal model of Cassinelli \&
Olson (1979) and Waldron (1984) was that the X-ray absorption was too large
to see any X-rays arising from regions near the photosphere.

Our primary target is the brightest X-ray source, Cyg OB2 No. 8a, and will
be the focus of our discussion, but some information about the other OB
stars will be presented. In Section 2 we discuss the data reduction and
present the HETG spectra of four \COB\ stars.  Our observational analyses
for obtaining line emission characteristics and the distribution of the
X-ray emission are presented in Section 3. The conclusions are discussed in
Section 4.

\section{The Chandra Observations of Cyg OB2}

\subsection{Data Reduction} 

Observations were taken with the \Chandra\ X-ray Observatory using the
ACIS-S CCD's with the HETGS in the optical path.  The observation of the
\COB\ association [Observation Identification Number (ObsID) 2572] began on
July 31, 2002 at 1$^h$ 51$^m$ 04$^s$ UT with an exposure time of
approximately 65.12 ks.  The spacecraft aim point was chosen to place the
primary target, cluster member Cyg OB2 No. 8a, in the center of the focal plane.
This gives us a sharp zeroth order image and the highest energy resolution
spectrum possible for this star.  During this observation, the spacecraft
aspect angle was chosen to allow recovery of the dispersed spectra of the
three other stars, Cyg OB2 Nos. 5, 9, 12. A ``true-color'' image constructed from the data
is displayed in Figure 1 which shows the typical rainbow colors of the High Energy Grating
(HEG) and Medium Energy Grating (MEG) dispersed spectra (e.g., Figure 8.1 in the Chandra
Proposers Observatory Guide, CXC 2002).

The CIAO tool TGDETECT was used to locate X-ray sources in the focal plane
image.  The sources found are displayed in the middle frame of Figure 1.  A
comparison with the Digital Sky survey image of the same region immediately
allows the four brightest sources to be identified with Cyg OB2 Nos. 8a, 5,
9, and 12.  Table 1 lists the adopted stellar parameters for these four
stars.  Although a number of other faint stellar sources are detected, this
paper concentrates on what can be learned from the four HETGS dispersed
spectra obtained during this observation, primarily Cyg OB2 No. 8a.  Since all
four stars have large ISM absorption column densities (see Table 1) the
extracted spectra reveal essentially no information for $\lambda > 12$ \AA.
The observed count rates for Cyg OB2 No. 8a are 0.102 (MEG$\pm$1) and 0.044
(HEG$\pm$1).  The only other spectrum strong enough to allow detailed
analyses of a few strong lines is Cyg OB2 No. 9 which has count rates of 0.022
(MEG$\pm$1) and 0.007 (HEG$\pm$1).

The intrinsic energy resolution of the ACIS-S CCDs act as
an effective cross dispersion, so individual photon events in the dispersed
spectra can be assigned relatively unambiguously to the correct source.
When extracting and calibrating the dispersed spectra for these four bright
point sources, the following issues needed to be kept in mind:

\begin{itemize}

\item Because these stars are off-axis, the High Resolution Mirror Assembly (HRMA) effective
area is somewhat less and the line-response function for Cyg OB2 No. 5, 9, and 12 is
somewhat broader (e.g., Figure 8.23 of the CPOG, CXC 2002).  In particular,
note the direct image of star Cyg OB2 No. 5.  It is so far off the optical
axis of the HRMA that its zeroth order
image is considerably broadened, greatly reducing the resolution of the
dispersed spectrum (See pp. 41-50 and 187-190 of CXC 2002).

\item As our primary target, the direct image of Cyg OB2 No. 8a is near
the center of the ACIS-I CCD array. Its whole spectrum for the HEG and MEG
fell on light-sensitive areas of ACIS-S.  Because of its location, this is
also the case for Cyg OB2 No. 5.  However, for stars Cyg OB2 Nos. 9 and 12, their
positions near the edge of the chip array means that part of their MEG -1
and HEG +1 spectra fell outside the light-sensitive area of the chip,
resulting in lower effective areas at longer wavelengths.

\item When dealing with an observation containing multiple X-ray sources,
such as this one, it is theoretically possible for an individual photon
event to be at the correct focal plane position and energy for more than one
source (Figure 8.28, CXC 2002).  If such a spatial-spectral ``collision''
occurs, the identification of a single source for that particular photon
would become ambiguous. This might occur when a MEG spectrum of one source
overlaps a HEG spectrum of another source (see the overlap between the
spectra of stars Cyg OB2 Nos. 8a and 5 in our observation).  To assess how
this problem might effect this observation, we independently constructed
photon event lists for the spectra of the four brightest X-ray sources in
the field: Cyg OB2 Nos. 5, 8a, 9, and 12.  We then examined the resulting
photon event lists, checking for any individual photon events which occurred
in more than one event list.  No such photons were found, indicating that
each of the extracted spectra are basically free from contamination by the
other sources in the field.  In each case, the observed similarity of the
positive and negative orders (when both were available) for each source also
confirmed that no such contamination occurred.

\item The direct images of Cyg OB2 Nos. 5 and 12 did not fall on
chip S3, the usual pointing direction.  The default chip set in the CIAO
tool MKGARF had to be reset to include chip S2 to prevent the erroneous
calculation of zero effective area for the portions of the HEG+1 and MEG+1
order spectra which fell on that chip.

\end{itemize}

High-resolution spectra derived from HETG-ACIS-S naturally have an extremely
low background because the intrinsic energy discrimination of the ACIS-S
effectively acts as a cross dispersion. This not only allows photon events
at a specific position along a spectrum to be sorted to the correct order,
but also causes most background events (which have incorrect energies for a
given location) to be rejected.

\subsection{HETG Spectra of the Cyg OB2 Stars} 

To survey the dispersed spectral data for the four bright stars, we co-added
the positive and negative first order spectra of each of the four stars (the
$2^{nd}$ and $3^{rd}$ order spectra were found to have too few counts to
make any significant contributions).  The HEG and MEG $\pm$1 count spectra
for Cyg OB2 No. 8a are shown in Figure 2, and Figure 3 shows the MEG $\pm$1
spectrum of Cyg OB2 No. 9.  Unlike other O-star spectra, the spectra of
these two stars are noticeably dominated by only two strong lines, the
H-like \SiXIV\ line and the He-like \SiXIII\ \fir\ lines. The source of this
difference from other O-stars is related to the greater high energy emission
associated with these stars, and the much larger ISM extinction.  In
addition, the Cyg OB2 No. 8a spectra reveals several higher energy ionic
line emissions (e.g., \ArXVII, \CaXIX, \FeXX, \FeXXII, \FeXXIII,
\FeXXIV, \&\ possibility \FeXXV).  Although \FeXX, \FeXXII, \FeXXIII,
and \FeXXIV\ were seen in our HETGS observation of \dori\ (Miller \etal\
2002), none of these ionic species were seen in our other O-star supergiant
HETGS observations.

The other two stars, Cyg OB2 Nos. 5 \& 12, have much weaker count spectra.
However, we can provide a comparison of all four stars by displaying their
associated MEG$\pm1$ flux spectra in Figure 4 which provides us with a first
order approximation of the X-ray line fluxes for comparisons among these
four stars.  For each star, a custom Ancillary Response File (ARF) and a Redistribution Matrix
File (RMF) were generated using the standard CIAO routines, and the flux spectra were obtained
by multiplying the count spectra by the bin energy and dividing by the associated ARF and
exposure time. Even though line-shape parameters cannot be recovered from
many lines in these spectra, the fact that individual lines are resolved in
the dispersed spectra still opens up avenues of investigation which are not
possible with CCD-resolution spectra such as a non-grating ACIS-S
observation.  However, as discussed in Section 2.1, artificial line
broadening can occur for off-axis sources due to variations in the
instrumental line spread function.  For example, some of the lines in the
spectrum of Cyg OB2 No. 5 appear much wider than seen in the other stars,
but this appearance is most likely caused by the large off-axis location of
the source.

\section{Observational Analysis}

\subsection{Measuring the X-ray Emission Line Properties \label{sec:lines}}

Since the X-ray line emission profiles
from OB stars are very broad ($HWHM~ \sim 1000$ \kms), early-type stars are among the few
classes of X-ray sources for which
line shapes can be resolved using the HETGS. This capability allows us to
study detailed line formation processes that we suspect are operating in
these stars. Although the currently accepted scenario is that a distribution
of stellar wind shocks are responsible for the observed X-ray emission,
\Chandra\ observational results are raising several key issues concerning the
exact nature of this distribution, including the possibility that not all of
the X-ray emission arises from the stellar wind.

A key parameter in determining the structure of an X-ray line profile from a
wind distribution of sources is the stellar wind column density scale factor
which is proportional to $\Mdot/(\vinf \Rstar)$ (see Waldron et al. 1998).
Since the mass loss rate of Cyg OB2 No. 8a is about 5 times greater than
\zpup\ (the only O-star showing well defined blue-shifted lines), we find
that this column density scale factor is $\sim 3$ times larger than for
\zpup.  Hence, well pronounced blue-shifts accompanied by highly asymmetric
line profiles are expected.

\subsubsection{Line Fitting Procedure \label{sec:linefit}}

We derive X-ray emission line properties using a relatively simple line
fitting procedure.  We assume that all emission lines can be represented by
Gaussian line profiles superimposed on an underlying bremsstrahlung
continuum.  The line emissivities and rest wavelengths are obtained from the
Astrophysical Plasma Emission Code and Database (APEC and APED; Smith \&
Brickhouse 2000).  For each line, the free parameters are the centroid shift
velocity ($V_S$), the $HWHM$ velocity, and the line normalization strength
(i.e., the line emission measure, $EM_X$).  Since we are only concerned with
the total flux within a line, we find that our fits are essentially
independent of our choice of the continuum temperature.  For all fits we
assume a continuum temperature of 10 MK. In this paper, we concentrate only
on the H-like and He-like lines. For each individual line or line complex,
we define a wavelength region large enough to cover the emission lines and
provide a good representation of the underlying continuum.  In many cases,
the defined wavelength region may include other weaker lines, and to ensure
we obtain the best estimate of the flux in the strongest line and continuum,
we include any lines which are likely to contribute to the observed
emission. When fitting a region with multiple lines, we always assume that
they all have the same $V_S$ and $HWHM$, and only the line and continuum
$EM_X$ will be different.  This approach is clearly justified for the
He-like \fir\ lines since all three of these lines must be formed under the
same conditions, and the only parameter that could be different is the
normalization of each line. Although all H-like lines and the He-like $i$
lines are doublets, their separations are much less than the resolving
capabilities of both the MEG and HEG, so the lines are represented by only
one Gaussian.

We use the standard $\chi^{2}$ statistics to determine a goodness of fit.
The quoted best-fit values for $V_S$ and $HWHM$ represent averages of their
respective $\chi^{2}$ 90\% confidence region range for each parameter, and
the quoted errors are associated with the difference from the average.
These errors do not include the associated MEG and HEG instrumental
wavelength uncertainties (e.g., see Table 8.1, CXC 2002).  The line flux
errors are determined by the total counting statistics within the line. Each
model line is folded through the HETGS instrumental response functions (ARF
and RMF) as determined for each star, and the best fit values are extracted.
To maintain consistency in fitting the MEG and HEG lines, the $\chi^{2}$
statistics are based on fitting a bin size of 0.01 \AA\ for both the MEG and
HEG which is approximately equal to the wavelength resolution limit of the
HEG instrument (0.012 \AA; CXC 2002).  Because the MEG and HEG have different line
response functions, we analyzed them separately.  It should be noted that the fits resulting from
the two instruments are consistent with one another, especially for the lines with the highest
signal-to-noise ratio (S/N).

\subsubsection{Determining the Degree of Line Asymmetry \label{sec:asymmetry}}

We test the degree of line asymmetry by analyzing the distribution of the
$\chi^{2}$ residuals. For example, suppose we have a triangular shaped
emission line (e.g., MacFarlane \etal\ 1991) with the peak of the triangle
located blue-ward from the rest wavelength and the line emission steadily
drops from this peak value to zero towards the red-ward side of the line
(there is no emission blue-ward of the peak).  By attempting to fit this
line with a model Gaussian line profile, we expect two characteristic
features to emerge: 1) the Gaussian fit will underestimate the line centroid
shift, and; 2) the residuals blue-ward of the Gaussian best fit line
centroid shift velocity will be positive (model underestimates the data),
whereas, red-ward from this centroid shift velocity, the residuals will be
negative (model overestimates the data).  Hence, a clear residual signature
is expected if there is asymmetry in an observed emission line.

To test this hypothesis, Figure 5 shows the Gaussian best-fit models for the
four strongest H-like emission lines of \zpup\, which is known to have large
blue-shifts and asymmetric line profiles (Cassinelli \etal\ 2001).  The
expected residual signature is clearly evident in the \FeXVII\ and
\OVIII\ lines, and possibility in the \NeX\ line.
The \MgXII\ line appears to have no line asymmetry based on our residual
analysis. Hence, we will use these \zpup\ line fit residual signatures as
our benchmark for establishing the degree of X-ray emission line asymmetries
in our Cyg OB2 No. 8a line profiles.  Note that this approach cannot be applied to regions of line
overlap such as \fir\ line complexes.

\subsubsection{Best-Fit Line Emission Characteristics \label{sec:linechar}}

We apply our line fitting procedure to the four H-like and four He-like
X-ray emission lines of Cyg OB2 No. 8a.  The results of our fits are given
Table 2 which lists the ion, rest wavelength, observed line flux, $V_S$, $HWHM$, $EM_X$, the
X-ray temperature (\TL) range where the line emissivity is within 75\% of its
expected maximum value, and the reduced $\chi^{2}$ .  The $EM_X$ are
derived as explained by Kahn \etal\ (2001), including ISM absorption
corrections, and their errors are based on the differences associated with
the range in emissivities as determined by the range in \TL.  In Table 2,
the fluxes for the He-like ion \fir\ lines represents the total flux from
all three lines, but the listed $EM_X$ corresponds to the $EM_X$ of the $r$
line only.  One has to be careful in their interpretation of $EM_X$ derived
from the $i$ and $f$ lines since the emissivities used to derive these
quantities have not been corrected for either density and/or UV effects.
Hence, the $r$ line $EM_X$ represents the most reasonable estimate since it
is essentially unaffected by these density and UV effects. The individual
fluxes from the three \fir\ lines are given in Table 3 (see \fir\ discussion
in Sec. 3.2).

The MEG and HEG H-like X-ray line fits and $\chi^{2}$ residuals are shown
respectively in Figure 6 and Figure 7.  Correspondingly, the MEG and HEG
He-like X-ray line fits are shown respectively in Figure 8 and Figure 9.
Although the line fits indicate a tendency for blue-shifted lines (see $V_S$
in Table 2), these shifts are smaller than those of \zpup, which have an
average blue-shift of $\sim -500$ \kms\ (Cassinelli \etal\ 2001).
Furthermore, these line shifts are significantly smaller than expected as
discussed in Section 3.1.1, and most of these line shifts are similar to
other O-star results which also show minimal or no blue-shifts.  However,
for a few lines, large blue-shifts are indicated (see Sec. 3.1.4). The
expected line broadness seen in O-star spectra is also present in the four
brightest Cyg OB2 OB stars.  Our detailed analyses of Cyg OB2 No. 8a
indicates that all H-like and He-like lines have $HWHM$ of $\sim
530$ to $1100$ \kms\ (neglecting upper limit values), with a suggestion, based
on the HEG fits, of a decrease at higher \TL. This behavior was also noted
by Waldron \& Cassinelli (2002) in their analysis of four O-stars.  However,
we note that even though a large range of \TL\ is represented in our sample
of lines, there is only a small relative change in $HWHM$.

From the resultant $\chi^{2}$ (Table 2) and visual inspection of the line
fits, we suggest that our assumed input Gaussian line profile model provides
reasonably good fits to the data. However, before we can claim that a
Gaussian profile is a realistic representation of the intrinsic line
profiles, we must also determine if there is any evidence of line
asymmetries in the observed line profiles.  By comparing our best-fit
$\chi^{2}$ residuals with the \zpup\ residuals (Fig. 5), we do not see any
residual signatures in the Cyg OB2 No. 8a lines that would suggest the
presence of line asymmetry.  However, there are two possible exceptions, the
MEG \SiXIV\ and \SXVI\ lines, where both lines show a departure from a
Gaussian in one wavelength bin (see discussion in Sec. 3.1.4).  Although we
stated that this residual test is not applicable to the He-like \fir\ lines,
we point out that there are no obvious residual patterns that could be
construed as evidence of line asymmetries with one possible exception.  The
MEG \SXV\ $r$ line shows evidence of model deficient counts in the blue
wings of the $r$ line, but again, this is not seen in the HEG \SXV\ line.
Hence, we suggest that since the majority of the Cyg OB2 No. 8a X-ray
emission lines do not show any evidence of line asymmetries, Gaussian model
line profiles are consistent with the observed line profiles.  This is
clearly contrary to our expectations as discussed in Section 3.1.1, and also
suggests that wind distributed X-ray source models (e.g., Owocki \& Cohen
2001) will have problems in trying to fit these line profiles unless the
mass loss rate is drastically reduced.  In fact, even for \zpup\, a
reduction in mass loss rate was also determined to be necessary in order to
explain the X-ray line profiles with a distributed X-ray source model
(Kramer \etal\ 2003).

\subsubsection{Comments on Peculiar Emission Line Characteristics 
\label{sec:linepeculiar}}

With regards to line profile shapes and line centroid shifts, our Cyg OB2 No. 8a results are not as
straightforward as those obtained from the other O-stars.  In
our earlier studies, we found that the X-ray emission line properties for a given star could be
categorized as either having asymmetric profiles with large blue-shifted lines, or symmetric
profiles with essentially no line shifts.  For example, we found that all emission lines in \zpup\
(Cassinelli \etal\ 2001) were found to have a large blue-shift of $\sim -500$ \kms\ with many lines
displaying asymmetric line profiles (see discussion in Sec. 3.1.2), whereas, for \dori\ (Miller \etal\
2002), all lines were found to have symmetric line profiles with a small range in shifts (i.e., $\pm
150$ \kms).  For Cyg OB2 No. 8a, we find a large range in the observed line centroid shifts
($V_S$ in Table 2), and no evidence for any line asymmetry (see Sec. 3.1.3).  Although it appears
that the Cyg OB2 No. 8a X-ray emission line properties do not fit either category, in the
following discussion we present arguments which suggest that the line properties of Cyg OB2 No.
8a may not be that unusual, and are in fact consistent with what we found for all other O-stars,
except for \zpup, in having lines that are symmetric with minimal blue-shifts.

From our analysis of the Cyg OB2 No. 8a X-ray emission lines, the range in line
centroid shifts is -832 to +245 \kms\ with no indication of any systematic trends in
these shifts (e.g., a temperature dependence).  However, since most of these large shifts are
associated with the lines from the higher energy ions (\SXV, \SXVI, \ArXVII) which have low
count rates, implying poorly determined shifts, we argue that these large shifts may not be real.  
If we ignore the line shifts from these higher ionization state lines, the resultant
range in centroid shifts reduces to -550 to +50 \kms\ which is still unusual as compared to other
O-star results.  The extremum of this range is due entirely to one line, the MEG \MgXII\ line with
a blue-shift of -550 \kms, and without this line, the range in line shifts reduces to a value
consistent with the majority of O-stars.  In fact, by averaging the MEG and HEG centroid shifts
of the strongest lines (\MgXI, \SiXIII, and \SiXIV, neglecting \MgXII), we find a mean shift of
only -91 \kms.  

The real problem is trying to understand the large blue-shift of the \MgXII\ line.  In particular, a
major discrepancy is evident when we compare the centroid shift of the \MgXII\ line with that of
the He-like \SiXIII\ line shift which shows a minimal blue-shift.  The discrepancy in these centroid
shifts is verified by both the MEG and HEG spectra.  
At first, the obvious explanation of this shift discrepancy is that the \MgXII\ line forms farther out
in the wind than the \SiXIII\ lines, but, based on our discussion in Section 3.1.1, the large centroid
shift seen in \MgXII\ should also be accompanied by a highly skewed line profile which is not
observed (see Sec. 3.1.3).  Regardless of this problem, the most puzzling aspect is that since both
of these lines have essentially the same ionic abundance dependence on temperature, and since
both of
these lines reach their maximum emission at $\sim$ 11 MK, whatever conditions are
responsible for forming the \MgXII\ line must also be forming the \SiXIII\ lines which implies that
both lines should have similar line properties.  Although both lines have symmetric line profiles,
the large difference in their centroid shifts is difficult to understand. 
We suggest that some of this discrepancy could be related to the low
S/N of the \MgXII\ line [$\sim$ 5 (MEG) and 4 (HEG)] as compared to the S/N of the \SiXIII\
lines [$\sim$ 9 (MEG) and 5 (HEG)], but, at this point, we conclude that the cause of the peculiar
blue-shift of the \MgXII\ line remains unclear. The resolution of this issue will require a higher
S/N observation.

In addition, the observed MEG \SiXIV\ and \SXVI\ line profiles display possible evidence for a
narrow highly blue-shifted strong emission component, which would suggest a departure from the
assumed Gaussian input line profile.  Although it is intriguing that both MEG lines show this
feature occurring at the same blue-shifted velocity of $\sim -700$ \kms\ (see Fig. 6), which is
significantly larger that the best-fit Gaussian line model predictions for these lines (see Table 2),
there are two issues that challenge the validity of this feature. First, the MEG \SXVI\ line is very
weak.  Second, this large blue-shifted component is not evident in either of the corresponding
HEG \SiXIV\ and \SXVI\ lines, but this could be related to the lower sensitivity of the HEG as
compared to the MEG. The HEG effective area in this spectral range is smaller than that of the
MEG by $\sim$ 44\% for \SiXIV\ and $\sim$ 58\% for \SXVI.  However, we do not wish to
ignore this feature completely since the shift might be real, and could be an indication of an
interesting new phenomenon occurring in early O-stars (e.g., a high velocity mass ejection from
some region of the stellar wind or possibility from the stellar surface).

\subsection{ Diagnosing the Location of the X-ray Emitting Plasma 
Using He-Like Ions \label{sec:fir}}

For OB stars the $f/i$ ratio provides information on the radial distance
from the star to the He-like ion line formation region (Kahn \etal\ 2001;
Waldron \& Cassinelli 2001).  Even though the He-like \fir\ lines often
overlap as demonstrated in Figure 10, line fitting procedures have been
successful in extracting the individual fluxes for each of the three \fir\
lines. From our studies of these He-like line formation regions, we have
been finding that for most OB stars, the high ion stages are formed with
\Rfir\ close to the star, which could indicate the presence of magnetically
confined regions on the surface (Waldron \& Cassinelli 2001).

It is important to point out that since the $f/i$ ratio is sensitive to the assumed photospheric flux
model, there is a model dependent uncertainty in the derived \Rfir (e.g., see discussion by Miller
\etal\ 2002).  This is particularly relevant for the \Rfir derived from the \ArXVII, \SXV, and
\SiXIII\ He-like $f/i$ ratios which are sensitive to the unmeasurable flux short-ward of 912 \AA. 
However, since this $f/i$ diagnostic has shown an overall consistent pattern in the radial locations
of the He-like X-ray emission lines for several O-stars, we believe that this diagnostic is currently
the most viable procedure for estimating the radial locations of the He-like X-ray emission lines,
with the understanding that the results are sensitive to the assumed photospheric flux model.

Following the procedure outlined by Waldron \& Cassinelli (2001) we have
determined the radial positions of the He-like ions of \MgXI, \SiXIII, \SXV,
and \ArXVII.  The MEG and HEG $f/i$ ratios and derived radial positions,
\Rfir, are given in Table 3. The $f/i$ ratio dependence on radius is
shown in Figure 11 where the MEG and HEG observed ranges in $f/i$ and their corresponding
radial ranges are indicated by darken curved sectors.  
Also shown is our standard
X-ray continuum optical depth unity plot as a function of radius which shows
that the \Rfir\ do appear consistent with other OB stellar results, i.e.,
the observed He-like X-ray emission is predominantly emerging from its
associated `effective' X-ray photosphere as first suggested by Waldron \&
Cassinelli (2001).  The calculations of these X-ray continuum optical depth
unity radii are discussed in the Appendix.  As evident in Figure 11,
although the mapping of \Rfir\ to the X-ray optical depth unity radii is not
exactly one-to-one, the overall mapping does appear to follow the wavelength
dependence of the optical depth unity radii reasonably well.  One notable
exception is that the \SXV\ \Rfir\ is significantly smaller than expected.
This could imply that the associated high temperatures required to produce
this emission are only produced deep in the wind ($< 1.1$ stellar radii)
and, hence, are actually located at a depth where the X-ray optical depth is
slightly $> 2$.  As seen in other O-stars, we also see that the higher ion
stages are progressively closer to the stellar surface and, in particular,
\ArXVII, is essentially on the surface (not shown in Figure 11).  Figure 11
also shows the wavelength dependencies of the X-ray continuum optical depth
unity radii for a mass loss rate 2 times smaller and 2 times larger than the
value given in Table 1. The best match between \Rfir\ and X-ray optical
depth unity radii appears to be associated with a mass loss rate that is
$\sim 2$ times smaller.  In a recent study, Hanson (2003) suggests that the
Cyg OB2 association may be $\sim 35 \%$ closer than originally though which
would imply a reduction in the radio determined mass loss rate (given in
Table 1) by a factor of $\sim 2$.

\subsection{Temperature Diagnostics of the X-ray Line Emitting Plasma}

Of particular interest regarding OB stars is the temperature of the X-ray
emitting plasma.  There can be several causes of this X-ray emission,
ranging from shocks embedded in winds, bow shock structures either around
clumps in winds or around unseen companion stars, or magnetic structures
near the base of the wind.  For example, determining the relationship
between $EM$ and $T$ could provide clues on the X-ray formation process.
Here we discuss two line ratio diagnostics techniques to
derive X-ray emitting plasma temperatures.

Miller \etal\ (2002) were the first to apply the H-like to He-like line
ratio temperature diagnostic to an early-type star, the late O-star \dori.  In general, one
might expect problems using this diagnostic for X-rays embedded in a dense
wind of an OB star since a small change in position could produce
significantly different wind attenuations. However, it is becoming
increasingly clear that each X-ray emission line appears to be arising from
its associated X-ray continuum optical depth unity position.  Hence, when
taking line ratios, the attenuation factors are roughly equal and cancel.  Normally,
the He line used in the $H/He$ line ratio is the sum of all \fir\ lines.
However, to avoid the uncertainties associated with the
formation of the $f$ and $i$ lines as discussed earlier, we only use
the He-like $r$ line in our calculation of the $H/He$ line ratio. We have used the APED data to
determine the dependence
of the $H/He$ ratio on temperature.  Table 4 lists the available H-like to
He-like line ratios and their derived X-ray temperatures, $T_{H/HE}$.
Notice that both the MEG and HEG derived temperatures are in very good
agreement. Since we expect that the two lines associated with each $H/He$
ratio are probably formed in different wind locations, our interpretation of
this temperature diagnostic is that it represents the average X-ray
temperature between these two locations and, 
in fact, all of these $T_{H/HE}$ are found to lie within the temperature range specified by their
respective H and He line \TL\ values.  
All of these temperatures along with their associated \Rfir\ support earlier claims that, in general,
the X-ray temperatures associated with the majority of the observed X-ray emission in OB stellar
winds are increasing inward toward the stellar surface.

Another line ratio technique used extensively in solar studies is the
He-like $G$-ratio [$G = (f+i)/r$].  Waldron \& Cassinelli (2001) were the
first to apply this technique to an early-type star, the O-star \zori, where they found good
agreement with \TL.  The observed ISM corrected $G$-ratios for our He-like
lines are given in Table 3, and the associated temperatures, $T_{fir}$, are
determined by comparing our observed $G$-ratios with $G$-ratios calculated
from the APED data.  Except for the HEG \MgXI\ $G$-ratio, there appears to
be significant problems with this method as compared with \TL\ and
$T_{H/He}$. One possible explanation is that the $r$ line flux may be
reduced by strong resonance line scattering which results in a higher
$G$-ratio and a smaller temperature (Porquet \& Dubau 2000).  However, there
are relatively large discrepancies in $G$-ratio values quoted throughout the
literature.  For example, the $G$-ratio temperature relation used in the
analysis by Waldron \& Cassinelli (2001) is found to be significantly
different than those quoted by Porquet \& Dubau (2000).  As these line ratio
discrepancies are resolved, the usefulness of this technique for OB stars
needs to be explored in greater detail.

\subsection{The Emission Measure Distribution}

The range of temperatures in the X-ray forming regions tells us about the
nature of the shocks involved. Cohen, Cassinelli, \& MacFarlane (1997) found that for the near
main sequence star \tsco, $EM(T) \propto T^{-2}$. Such a power law result is
also found for bow shock models of clump generated X-rays from
hydrodynamical theory (Moeckel, Cho \& Cassinelli 2002). In the Moeckel \etal\ bow shock
model, it was found that a wide range of ionization conditions
could be present both because there can be very hot matter right at the peak
of the bow and a whole range of cooler material produced at the oblique
shock region of the bow. The detection of \ArXVII\ line emission, for example, indicates that
there is a hotter source of gas in Cyg OB2 No. 8a than in other O-stars we have
studied, such as \zpup.  Moeckel \etal\ (2002) showed that the emission
measure versus $T$ distribution behind a bow shock is a power law where
$EM(T) \propto T^{-4/3}$. The modeling of \tsco\ by Howk \etal\ (2000) indicated
that the wind is strongly influenced by bow shocks. Miller (2002)
found a rough power law dependence for \dori\ with a slope of $-2/3$ and it
was argued that there could be a wind collision with a companion
star. However, based on the $EM$ values listed in Table 2, there is no clear
indication of any temperature dependence and, in fact, one could argue that
all line $EM$ values are essentially the same.

\subsection{Spectral Fit to the HETG/MEG Spectrum of Cyg No. 8a}

In general, fitting the overall HETG spectra of OB stars is a difficult task
due to the large range in temperatures, the distribution of X-ray sources,
the widely varying degree of stellar wind absorption throughout the wind,
the extreme line broadening, and the quenching of the He-like $f$ line.  For
the case of Cyg OB2 No. 8a, since there is no observed soft X-ray emission
due to the large degree of ISM extinction, a model with fewer X-ray source
temperatures can be used. Hence, our goal here is to find the 
simplest  model capable of reproducing the overall spectral characteristics
observed in the MEG spectrum of Cyg OB2 No. 8a.  In addition, this model fit
provides us with the total observed X-ray flux in the HETGS energy band
width which can be used for comparisons with other observed broad-band X-ray
results.  We use the \ROSAT\ (Waldron \etal\ 1998) and
\ASCA\ (Kitamoto \& Mukai 1996) derived fitting parameters as starting
points for establishing temperature and column density estimates.  The X-ray
emission lines and continuum are calculated using the MEKAL emissivity model
(Mewe, Gronenschild, \& van den Oord 1985).  In addition, to account for the large line
broadening
and peculiar He-like \fir\ line behavior, the emissivity model had to be
modified by considering: 1) all emission lines are assumed to be Gaussian in
shape and include a pseudo ``turbulent velocity'' component to mimic the
line broadening (which is reasonable considering that all lines are
generally symmetric), and; 2) artificially large X-ray densities are
included to suppress the He-like forbidden line emission since the current
model has no provisions for modeling the effects of an ambient UV radiation
field. The model fits are then obtained by folding the spectral model
through the HETGS instrumental response functions (RMF \& ARF).  The free
parameters are the temperatures, wind column densities, and emission
measures.  The fixed parameters are the ISM column density given in Table 1,
a turbulent velocity of 1150 \kms, and for the low and high temperature
components, the emissivity densities are respectively $3.2 \times 10^{13}$
and $1.0 \times 10^{13}$ cm$^{-3}$.  Note, these densities have no physical
meaning with regards to the X-ray emitting plasma since they are only used
to simulate the correct $f/i$ ratios which are controlled by the stellar
UV/EUV radiation field.

We find that a two-temperature, two-wind column density model
can provide an adequate overall fit to the MEG spectrum.  The resultant
temperatures are $3.98\pm 1.67$ and $13.88\pm 1.80$ MK.  Their respective
wind column densities are $(1.22\pm 0.60) \times 10^{22}$ and $(1.41\pm
0.45) \times 10^{22}$ cm$^{-2}$, and their respective intrinsic emission
measures are $(3.65\pm 2.41) \times 10^{57}$ and $(1.06\pm 0.24) \times
10^{57}$ cm$^{-3}$.  The reduced $\chi^{2}$ is 1.29 ($\chi^{2}/DOF$ =
605/470).  Comparing these emission measures with the ISM corrected line
emission measures at similar temperatures (see Table 2) indicate that these
emission measures are approximately 17 and 4 times larger, and the
difference is related to the fact that these model derived emission measures
represent their intrinsic values (i.e., the $EM_X$ in Table 2 have not been
corrected for wind absorption). These two temperatures and wind column
densities are consistent with the \ROSAT\ and \ASCA\ derived parameters, as
well as the observed log \LxLbol\ of $\sim -6.75$.  Although the \ROSAT\ and
\ASCA\ fits only required one column density, we found that two wind column
densities were required in order to fit the relative strengths of the
\MgXII\ line with respect to the \MgXI\ and \SiXIII\ lines.  The majority of
the \MgXI\ emission is primarily associated with the 3.98 MK component
whereas, the majority of the \MgXII\ and \SiXIII\ line emissions are
associated with the 13.88 MK component. Furthermore, these best fit column
densities can be used to estimate the location of the two temperature
components assuming a spherically symmetric wind (see Waldron \etal\ 1998).
The resultant radius for the 3.98 MK component is $6.44\pm 2.97$ \Rstar, and
for the 13.88 MK component, the radius is $4.79\pm 1.38$ \Rstar.  Although
the associated errors in these radial positions indicate that these
locations are essentially indistinguishable, these model best-fit radial
locations do suggest that the hotter component is located slightly deeper in
the wind than the cooler component which is consistent with our \fir\
analysis (see Sec. 3.2).  In addition, the cool component location is found
to fall within the combined MEG and HEG \MgXI\ \Rfir\ range of 2.7 to 6.1
\Rstar (see Table 3), and is also consistent with the single component model
fit to \ROSAT\ data which predicted an X-ray location of $\sim 4.7$ \Rstar\
(Waldron \etal\ 1998). The hot component location is found to be larger than
the combined MEG and HEG \SiXIII\ \Rfir\ range of 1.7 to 2.4 \Rstar.  The
most likely explanation of this discrepancy is the simplicity of assuming a
two-component fit model. For example, by considering a model with a
distribution of X-ray temperatures around the current hot component
temperature, significantly more emission over a broader spectral energy
range would be produced.  This excess emission would have to be balanced by
an increase in the absorbing wind column density, which in turn would
predict a smaller radial location range for this distribution of
temperatures.  In principle this same argument should also be applicable to
the cooler component.  However, since most of this soft X-ray emission is
masked by the large ISM absorption, and this observed soft emission has very
weak lines with essential no continuum, it is understandable why a single
temperature component provides a reasonable fit to the softer spectral
region.

The comparison of our model fit with the observed spectrum is shown in
Figure 12 which only covers the wavelength region of the strongest lines.
This shows that the line strengths, broadness, and $f/i$ ratios are in very
good agreement. The only real discrepancy is in the region between 7 and 8
\AA\ where the model has a problem in fitting these weaker lines. The
\Chandra\ observed log \LxLbol\ is found to be $\sim -6.75$ which is
essentially identical to the value determined when last observed in 1993
(within $\sim$ 20 \%).  Waldron \etal\ (1998) studied the long term X-ray
variability of these Cyg OB2 stars and noticed that, of the four, the
observed X-ray emission from Cyg OB2 No. 8a had remained essentially
constant for $\sim 15$ years.  We can now suggest that Cyg OB2 No. 8a X-ray
emission has now remained constant for the last 24 years.  In addition,
since the \LxLbol\ ratio is consistent with the general O-star behavior, we
can argue that the X-ray emission processes in Cyg OB2 No. 8a are probably
no different than those occurring in isolated O-stars. In particular,
Chlebowski (1989) found that the \LxLbol\ ratio for close binary stars have
significantly larger values than the observed \LxLbol\ value for Cyg OB2 No.
8a.

\subsection{Analysis of the Cyg OB2 No. 9 MEG Spectrum}

As summarized by Waldron \etal\ (1998), Cyg OB2 No. 9 is by far one of the
most interesting variable stellar radio sources where essentially every time
a radio observation is obtained it displays a different structure.  For
example, it has displayed both thermal and non-thermal characteristics at
different epochs.  However, even when it does appear in a thermal state, its
radio emission is still different, which, assuming free-free emission,
indicates a highly variable mass loss rate.  On the other hand, with regards
to X-ray emission, Cyg OB2 No. 9 is similar to Cyg OB2 No. 8a in that the
X-ray emission has remained relatively stable for $\sim 15$ years.  In
addition, Cyg OB2 No. 9 is also the weakest of the four in terms of X-ray
emission and the strongest radio emitter, whereas, Cyg OB2 No. 8a is the
strongest X-ray source and the weakest radio source.

Although Cyg OB2 No. 9 is the weakest of the four main stellar X-ray sources
in Cyg OB2, it is the closest of the remaining three to the aim point (Cyg
OB2 No. 8a), resulting in the second best dispersed spectrum observed.
However, the HETG spectrum of Cyg OB2 No. 9 is weak, so detailed analyses
like those for Cyg OB2 No. 8a are not possible.  Also, the bigger
instrumental line spread function makes the emission lines of Cyg OB2 No. 9
more difficult to measure.  Nevertheless by using an RMF-based analysis to
study these weak emission lines, reasonable reconstructions of line shapes
can be obtained. The MEG \SiXIII\ and \SiXIV\ lines are strong enough to
extract information on line profile parameters and line fluxes, and
reasonable line flux estimates can also be extracted for the \MgXI\ ($r$
line only) and \MgXII\ lines.  The following flux values have units of ergs
cm$^{- 2}$ s$^{-1}$.  For \SiXIV\ we find a flux = $0.17 \pm0.03 \times
10^{-13}$, V$_S$ = $-100 \pm270$ \kms, and a $HWHM < 940$ \kms.  For
\SiXIII\ we find a total \fir\ flux$ = 0.48
\pm0.1 \times 10^{-13}$, V$_S$ = $-700 \pm250$ \kms, and a
$HWHM = 620 \pm 430$ \kms.  The individual \fir\ line fluxes in units of
$10^{-13}$ are respectively $0.12\pm0.02$, $0.19\pm0.03$, and $0.16\pm0.3$,
which yield an $f/i$ ratio of $0.60 \pm0.16$. Correspondingly, the \MgXI\ $r$
line flux = $0.11\pm0.02 \times 10^{-13}$, and the \MgXII\ flux =
$0.08\pm0.02 \times 10^{-13}$.

From this limited amount of information, we find three interesting results.
First, the H-like to He-like line ratios for Mg and Si (see Table 4) provide
temperatures that are in very good agreement with the $T_{H/He}$ derived for
Cyg OB2 No. 8a.  Second, the \SiXIII\ $f/i$ ratio suggests a radial location
range between 1.4 to 1.9 stellar radii.  As discussed above, Cyg OB2 No. 9
had been observed twice when it had a thermal radio spectrum.  The
associated mass loss rate estimates from these two times are $12.7 \times
10^{-6}$ \Msunyr\ (1983) and $40.0 \times 10^{-6}$ \Msunyr\ (1993). Using
the 1983 mass loss rate value for determining the X-ray continuum optical
depth unity radius for \SiXIII\, it is found to be $\sim 1.5$ stellar radii.
The 1993 value predicts an X-ray continuum optical depth unity radii of
$\sim 4$ stellar radii.  Since the 1983 value provides a consistent radius
with the $f/i$ radius we suggest that the 1983 determined mass loss rate is
a better estimate of the mass loss rate for Cyg OB2 No. 9 at the time of our
observations.  Third, the fit to the \SiXIII\ lines suggests a rather large
blue-shift of $\sim -700$ \kms. If correct the implications are very
interesting, for we know from the $f/i$ ratio that the radial location of
\SiXIII\ is between 1.4 to 1.9 stellar radii. These radii correspond to a wind
velocity range of 915 to 1200 \kms\ (using the standard velocity law with a $\beta = 0.8$;
Groenewegen, Lamers, \& Pauldrach 1989). By assuming a saw-tooth wind shock structure as
proposed by
Lucy \& White (1980), this wind velocity range implies a shock velocity
range of 215 to 510 \kms\ and a corresponding shock temperature range of
only 0.6 to 3.6 MK, well below the expected temperature needed for formation of
\SiXIII\ ($\sim 11$ MK).  We also see a similar situation from
the \SiXIV\ line and \SXVI\ lines observed in Cyg OB2 No. 8a (see Table 2
and discussion in Sec. 3.1.4).  Thus it is difficult to understand these
observations in the context of this relatively simple shock model.

The line-spread functions of the other two stars (Cyg OB2 Nos. 5 \& 12) are
too large to be useful for extracting line shape information and $fir$
analyses which are the focus of this paper.  However, their MEG flux spectra
are included in Figure 4 which allows one to extract reasonable line flux
estimates for the various observed emission lines and provides a comparison
of the energy flux spectra of all four stars.

\section{Conclusions:  Are the X-ray Properties of Cluster 
Stars Similar to Those of Normal OB Stars?}

There are several reasons why the \COB\ region is of special interest.  The
strongest X-ray source in this region, Cyg OB2 No. 8a, is the second example
of a very luminous early O-star and it is one of the very few that \Chandra\
will be able to observe within a reasonable exposure time. Since we are
observing a cluster, our one observation provided high resolution spectra of
three early O-stars and one B supergiant. These objects could be of special
interest because Cyg OB2 might be a young globular cluster. Thus it might be
like the X-ray emitting regions observed at lower spectral resolution in
other galaxies.

The Cyg OB2 stars are in a relatively tight cluster that has been called a
young globular cluster (Kn\"{o}dlseder 2000; Hanson 2003). Hence, it is
possible, maybe even likely that there is a contribution to the X-ray flux
from all of the stars from wind-wind or wind-star collisions. For example,
Bonnell \& Bate (2002) found that star formation simulations indicate that
massive stars are generally in binary systems which can be relatively wide.
Thus while it is likely that $> 1/2$ of all OB stars are in binary systems,
these could be too widely separated for a predominance of X-rays to be
arising from wind-wind collisions.  Now if Cyg OB2 No. 8a were to have its
X-rays forming from such a collision, then we would see evidence for this
in our wavelength versus radius plots. That is, instead of having the radii
of line formation located near X-ray optical depth unity, these radii would
be much larger, comparable to that of the separation of early-type stars.
A hint of such a departure was seen in our study of \dori\ (Miller \etal\
2002), where \SiXIII\ was found to be formed well beyond the X-ray optical
 depth unity radius.  However, for Cyg OB2 No. 8a, since the lines do seem
to be formed near X-ray optical depth unity, we do not see a need for
postulating anything other than this star being similar to other early-type
single stars.

All previous X-ray observations (\Einstein, \ROSAT, \& \ASCA) of the
brightest Cyg OB2 Association OB stars indicate that these stars show clear
evidence that the X-ray emission is significantly hotter than most other OB
stars. This is also evident from our \Chandra\ observations where we clearly
see \ArXVII\ and other higher energy ions. From this feature along with the
fact that these stars are very massive, many have suggested that these stars
are precursors to Wolf-Rayet star formation (e.g., Schaller \etal\ 1992).
Unfortunately, due to the very large ISM extinction for the Cyg OB2 stars,
it is very hard to distinguish whether the ISM or wind is the dominant
absorbing component for the soft X-ray emission as also noted by Waldron \etal\ (1998).  Based
on our two component fit to the Cyg OB2 No. 8a MEG
spectrum, we find the wind column densities required to fit the spectrum
need to be larger than the ISM column density, suggesting that the wind
column density is dominant.  However, if these are precursors to WR stars,
then why do they have similar standard O-star behavior (e.g., comparable
\LxLbol\, and consistency with X-ray optical depth unity)?  Ignace, Oskinova, \& Brown (2003)
point out that X-ray optical depth unity radii are expected to be in a range
of a 100 to 1000 stellar radii for WR stars. One main difference between
an O supergiant and a WR star is the stellar radius, and as discussed in
Section 3.1, the wind column density scale factor is inversely proportional
to the stellar radius.  For example, assuming that the mass loss rate and
wind speed remains the same and the stellar radius of Cyg OB2 No. 8a is
reduced by a factor of 3 (from 30 to 10), we find that the location of X-ray
optical depth unity for 15 \AA\ changes from $\sim$ 10 to 50 stellar
radii. Furthermore, we know that WR star mass loss rates are typically
larger than those of the Cyg OB2 stars which would raise this X-ray optical
depth unity radius up to several hundred radii, consistent with WR results.

Although these high X-ray temperatures inferred from our \Chandra\
observation of the Cyg OB2 stars are unusual for O supergiants, they are not
overly unusual with regards to all O-stars.  For example, the \Chandra\
HETGS observations of the main sequence O5 star, \toriC, shows evidence for
even higher temperatures and is believed to be related to magnetic activity
(Schulz \etal\ 2003).  In addition, recent XMM-Newton observations of the
late O supergiant \zori\ (Mewe, private communication) shows evidence for
X-ray emission from \ArXVII\ which was not seen in the HETGS observations of
this star (Waldron \& Cassinelli 2001).  This is interesting since
it may be indicating the presence of a variable high X-ray energy component,
possibility associated with surface activity.  A \ROSAT\ time series
observation of \zori\ also indicated the possibility that some sort of
`flare-like' event was observed (\Berghofer\ \& Schmitt 1994).

To summarize with regards to our detailed analysis of Cyg OB2 No. 8a, we do
not see evidence for the strong wind absorption like that seen in WR stars.
We also do not see any strong evidence supporting wind-wind interactions as
evident from our \fir\ analysis and the derived \LxLbol. Thus we suggest at
this point to treat this star as if it is a single O-star X-ray source.
Using the single star approach we have come to these interesting
conclusions.

a) Two methods were used to derive the radius of X-ray formation, the $f/i$
ratio of He-like ions, and a fitting of the entire spectrum with a two
component model with temperatures and column densities for each source as
adjustable parameters.  Both methods are found to predict that the overall
radial range of the \MgXI\ and \SiXIII\ line formation region is 1.7 to 6.1 stellar radii.

b) Although this radial range is far below the range expected for X-rays
from WR stars, this does not rule out the possibility that this star is a
proto-WR star.

c) X-rays are forming deep in the wind, as was the case for \zpup\ and
\zori\, and essentially all of the \Chandra\ observed OB stars as shown in the
temperature versus radius figure of Cassinelli, Waldron, \& Miller (2003).
This demonstrates that the simple idea that the X-rays arise from wind
shocks of about $1/2$ the local wind speed is not correct. The cause of the
hot gas near the stellar surface thus remains a puzzle.

d) Because the Cyg OB2 No 8a and \zpup\ are similar early O-stars with
strong winds, we had expected that Cyg OB2 No. 8a would be more like \zpup\
in showing skewed and blue-shifted line profiles, but that is not the case.
Thus \zpup\ still appears to be unique among O-stars. However, it also means
that there is still a problem in understanding of how OB stars can have
symmetric X-ray line profiles. There are several explanations: 1) the winds
are so porous owing to fragmentation (e.g., Feldmeier \etal\ 2003) that one
can see farther into the wind, even to the far side of the wind where
red-shifting of line occurs; 2) the X-rays are arising from outflowing and
in-falling clumps (Howk \etal\ 2000), but their model should only be valid
for stars with very weak winds that allow for stalling and inflow; 3) Ignace
\& Gayley (2002) suggest that resonance line scattering may reduce the
degree of line asymmetry but their use of a Sobolev analysis and its
dependence on a smooth velocity gradient seems to be a questionable
treatment of an X-ray shock region, and; 4) the X-rays are forming in
extended magnetic loops, as had originally been envisioned by Underhill
(1980) and we see both hot up flows and down flows in magnetic tubes.

e) We were able to get X-ray information regarding four O-stars with one
observation that had a carefully planned satellite orientation. Although we
were only able to get the most detailed information from the primary source,
Cyg OB2 No. 8a, we demonstrated that we were also able to get some useful
emission line information from one other source, Cyg OB2 No. 9, and find
some reasonable line flux estimates for the two weakest sources (Cyg
OB2 Nos. 5 \& 12).

Because the OB stars considered here are in a tight stellar cluster, we have
had to consider whether the X-rays were dominated by collisions that produce
bow shock structures or by processes involving enhance magnetic confinement
of hot plasma. The arguments that we had to consider should be of interest
to X-ray astronomers studying galaxies and star burst regions, since they
are surely looking at clusters of OB stars somewhat analogous to Cyg OB2. In
our analysis we have developed criteria for deciding on the nature of the
X-ray emission.  In the case of Cyg OB2 No. 8a, we found no evidence that its X-ray emission is 
dominated by processes other than those occurring in isolated OB stars.

WLW was supported by NASA grant GO2-3028C and acknowledges partial support
from NASA grant GO2-3027A.  JPC and NAM were supported by NASA grant
GO2-3028.  NAM and JCR acknowledge support from the UW-Eau Claire Office of
Research and Sponsored Programs.  We wish to thank Shantih Spanton for some
initial discussions of the data.

\appendix
\section{Calculation of X-ray Continuum Optical Depth Unity}

Analyses of the He-like ion $f/i$ line ratios in OB stars suggest that the
derived $fir$-inferred radii are highly correlated with their respective
X-ray continuum optical depth unity radii (i.e., the radial wind location
where the X-ray continuum optical depth, $\tau_{\lambda}$ = 1, for a given
wavelength).  In general, the stellar wind X-ray continuum optical depth for
a given wavelength, $\lambda$, measured outward from radius $r$ is defined
as

\begin{equation}
\tau_{\lambda}(r)  = \int_{r}^{\infty}  \sigma_{\lambda}(r')  n_H (r') dr'
\end{equation}
where $\sigma_{\lambda}(r)$ is the radial and wavelength dependent wind
absorption cross section (cm$^{2}$), and $n_{H}$ is the hydrogen number
density (cm$^{-3}$) defined in terms of the wind mass density as $n_{H}(r) =
\rho (r)/\mu_{H}m_{H}$ (where $m_{H}$ is the atomic weight of hydrogen and
$\mu_{H}$ is the mean atomic weight).  Although we follow the basic
procedure for determining $\sigma_{\lambda}(r)$ as discussed by Waldron
(1984), several modifications have been incorporated.  In particular, we
have updated the photoionization cross sections (including all K-shell cross
sections from all ion stages), the collisional and recombination rates
(including dielectronic recombination), and all other relevant atomic data
(e.g., elemental abundances) that are available in the most recent version
of the Raymond \& Smith (1979) emissivity code as summarized by Raymond
(1988).  In addition, the calculation of $\sigma_{\lambda}(r)$ now includes
the contributions from 13 elements (H, He, C, N, O, Ne, Mg, Si, S, Ar, Ca,
Fe, Ni) and all their stages of ionization.

To obtain an analytic relationship between radius and $\tau_{\lambda}$, two
basic assumptions are required.  First, it is necessary to simplify the
integration in eq. (A1).  This is accomplished by assuming that
$\sigma_{\lambda}(r)$ can be represented as a density weighted average cross
section throughout the wind which is only dependent on the wavelength and
independent of the radial position.  This density weighted cross section is
defined as

\begin{equation}
\sigma_{avg}(\lambda) ~ = ~ \frac{\tau_{\lambda}(\Rstar)}{N_W (\Rstar)}
\end{equation}
where $\tau_{\lambda}(\Rstar)$ and $N_{W}(\Rstar)$ are respectively the
total X-ray continuum optical depth and wind column density (cm$^{-2}$) as
measured from the stellar surface.  A representative energy dependent
distribution of $\sigma_{avg}(\lambda)$ for the Cyg OB2 O- stars is shown in
Waldron \etal\ (1998).  Hence, $\tau_{\lambda}$ can be approximated as

\begin{equation}
\tau_{\lambda}(r) ~ \approx ~ \sigma_{avg}(\lambda)N_{W}(r) ~ 
= ~\sigma_{avg}(\lambda) \int_{r}^{\infty} n_H (r') dr'
\end{equation}
where $N_{W}(r)$ represents the wind column density measure outward from the
radial position $r$.  The error introduced by using the approximate
$\tau_{\lambda}$ (eq. A3) as opposed to the actual $\tau_{\lambda}$ (eq. A1)
is found to be only a few percent.  The second assumption requires
specification of the wind geometry and wind velocity law.  We adopt a
spherically symmetric wind with a wind speed that is determined by the
standard $\beta$ velocity law [i.e., $v(r) = \vinf (1 - \Rstar/r)^{\beta}$].
Although this velocity law allows us to obtain a simple analytic expression,
it produces a singularity in the wind density at the stellar surface.  In
the actual calculations used to determine $\sigma_{avg}(\lambda)$ and
produce the plots shown in Figure 11, the velocity law is modified by adding
a small but finite constant velocity term ($V_{\circ}$) to the standard velocity
law, and $N_{W}$ has to be determined numerically.  The addition of this
constant term places an upper limit to the wind density at \Rstar, and the
value of $V_{\circ}$ is typically chosen to be equal to approximately one-half
of the photospheric sound speed.  We realize that for the special case of
$\beta = 1$, an analytic expression is still obtainable with the addition of
$V_{\circ}$, but here we will adopt the more appropriate value of $\beta = 0.8$
(Groenewegen \etal\ 1989).  With these assumptions, the value
of $N_{W}(r)$ is given by (for $\beta~\neq~1$)

\begin{equation}
N_{W}(r) ~ = ~  \int_{r}^{\infty} n_H (r') dr' ~ = ~ {\frac{N_O}{1 - \beta}}
[1~ - ~ w(r)^{1/(\beta - 1)}]
\end{equation}
where

\begin{equation}
N_O  =  \frac{\Mdot}{4\pi \mu_{H} m_{H}\vinf \Rstar}~~~{\rm cm}^{-2}
\end{equation}
and $w(r) = v(r)/\vinf$. Although our expression for $N_{W}(r)$ given by eq.
(A4) is valid for essentially the whole wind structure, it will begin to
overestimate $N_W$ when $v(r)$ becomes comparable to $V_O$ which occurs at
radii below $\sim 1.01 \Rstar$.  Below this radius limit the overestimate in
$N_W$ is initially a few percent, reaching a maximum of $\sim 25\%$ at
\Rstar. However, the details in this region are immaterial to our discussion
concerning the location of X- ray optical depth unity since any depth
indicating a location $< 1.01 \Rstar$ can be interpreted as occurring
essentially on the stellar surface.  For additional details, and the case
for $\beta = 1$ see Waldron \etal\ (1998).

Since our goal is to find the relationship between radial position and
wavelength for a given value of $\tau_{\lambda}$ (e.g.,$\tau_{\lambda}$), we
proceed by first finding an expression for $w(r)$.  By combining the results
of eqs. (A3) and (A4) we have

\begin{equation}
w(r) =  \left[ 1 -  {\frac{(1 -\beta)\tau_{\lambda}(r)}{\sigma_{avg}(\lambda)
N_{\circ}}} \right]^{\beta/(1-\beta)}
\end{equation}
Using the adopted velocity law to obtain the relation between $r$ and $w$,
our desired relationship between radius and wavelength for a given
$\tau_{\lambda}$ is given by

\begin{equation}
r ~ = ~ \Rstar~\left( 1~ - \left[1~ - ~ {\frac{(1 - \beta)\tau_{\lambda}}
{\sigma_{avg} (\lambda) N_{\circ}}} \right] ^{1/(\beta-1)} \right)^{-1} 
\end{equation}

\clearpage

\onecolumn

\begin{table}[t]
\caption{Adopted Stellar Parameters \tablenotemark{a} 
\label{tab:stellar_properties}}
\begin{tabular}{lcccc}
\tableline
\tableline
     & Cyg OB2 No. 9  & Cyg OB2 No. 8a & Cyg OB2 No. 5  &  Cyg OB2
No. 12 \\ 
\tableline
Spectral Type & O5f   & O5.5 I(f)\tablenotemark{b} & O6f+O7f  & B8 Ia    \\
$T_{eff}$ (K) & 44700 & 38500\tablenotemark{b}     & 39800    & 11200     \\ 
log $L_{bol}$(ergs/s)& 40.19  & 39.77\tablenotemark{b}  & 40.01 & 39.79\\
 $M$   (\Msun)& 160   & 90.5\tablenotemark{b}      & 118       &  71       \\ 
 $R$  (\Rsun) & 34    & 27.9\tablenotemark{b}      & 34        & 338       \\
\vinf (\kms)  & 2200  & 2650\tablenotemark{b}      & 2200      & 1400  \\  
 $d$ (kpc)    & 1.82  & 1.82           & 1.82      & 1.82       \\
log $N_{ISM}$(cm$^{-2})$) & 22.06\tablenotemark{d} & 21.92\tablenotemark{d} &
22.02\tablenotemark{d} & 22.23\tablenotemark{d}  
  \\
 \vsini (\kms)  & 145            & 95\tablenotemark{b}  & 180   & 75     \\ 
\Mdot (\Msunyr)& $12.7 \times 10^{-6}$ \tablenotemark{c} & $13.5\times10^{-6}$
\tablenotemark{b}      
& $34.5 \times 10^{-6}$ \tablenotemark{c}& $38.5\times 10^{-6}$ 
\tablenotemark{c}                \\
\tableline
\end{tabular}
\tablenotetext{a}{Table values from Bieging \etal\ (1989), unless marked otherwise}
\tablenotetext{b}{Table values from Herrero \etal\ (2002)} 
\tablenotetext{c}{Table values from Waldron \etal\ (1998) - Adopted mass loss rate for Cyg OB2
No. 9 based on our X-ray analysis discussed in Section 3.6} 
\tablenotetext{d}{Interstellar column density estimates derived from the 
observed E(B-V) values
of Abbott \etal\ (1984) using the relation $N_{ISM}/E(B-V)$  = $5.2 \times
10^{21}$ cm$^{-2}$ (Shull \& Van Steenberg 1985)}
\end{table}

\clearpage

\begin{table}[t]
\caption{X-ray Emission Line Properties for \cygate 
\label{tab:lines}}
\vspace{-0.1cm}
\begin{center}
\begin{tabular}{lccrcccc}
\tableline
\tableline
Ion & $\lamz$ & \linflux & V$_S~~~$ & $HWHM$& $EM_x$ &\TL\tablenotemark{a} &
$\chi^{2}/DOF$ \\
& \AA &  erg cm$^{-2}$ s$^{-1}$& \kms & \kms& $10^{56}$ cm$^{-3}$ & (MK) & \\
\tableline
 \multicolumn{8}{l}{\textit{\underline{Medium Energy Grating}} } \\   
\NeX & 12.132  & $0.27 \pm0.04$  &   $0\pm450$  & $651\pm468$ &
$1.36\pm0.50$   & $6.82\pm2.09$ & $37.3/46$ \\
\MgXI ($fir$) & 9.169 & $0.64 \pm0.07$  & $-191\pm380$  & $998\pm182$ &
$2.13\pm0.68$ & $6.71\pm10.70$ & $41.0/42$ \\
\MgXII &  8.419  & $0.44 \pm0.04$  & $-550\pm314$   & $1025\pm421$ &
$1.66\pm0.49$  & $11.45\pm 3.51$ & $16.2/19$ \\
\SiXIII ($fir$)  & 6.648   & $2.20 \pm0.16$  & $-87\pm151$  & $884\pm96$  &
$2.90\pm0.88$  & $11.03\pm3.09$ & $52.1/40$  \\
\SiXIV  & 6.180   & $1.08 \pm0.07$  & $-36\pm216$   & $1096\pm316$  &
$2.56\pm0.70$  & $18.85\pm6.26$ & $17.7/14$ \\
\SXV ($fir$)  & 5.039   & $1.37\pm0.27$  & $-195\pm640$   & $<830$  &
$2.23\pm1.02$  & $16.18\pm 4.96$ & $28.3/24$ \\
\SXVI  & 4.727   & $0.36\pm0.08 $  & $245\pm307$   & $986\pm380$  &
$1.42\pm0.55$  & $29.32\pm10.49$ & $10.8/10$ \\
\ArXVII ($fir$)  & 3.949   & $0.30\pm0.10$  & $-100\pm250$   & $<150$ &
$1.56\pm1.14$  & $23.74\pm7.89$ & $24.6/22$ \\
\multicolumn{8}{l}{\textit{\underline{High Energy Grating}} } \\      
\NeX         & 12.132  & $0.15\pm0.05$  & $-300\pm300$  & $<1400$ &
$0.76\pm0.44$  & $6.82\pm2.09$ & $22.2/46$ \\
\MgXI ($fir$) &  9.169  & $0.91\pm0.16$  & $-146\pm610$   & $910\pm389$ &
$3.64 \pm1.35$  &  $6.71\pm 1.70$ & $34.5/42$ \\
\MgXII &  8.419  & $0.59 \pm0.07$  & $-432\pm283$  & $1019\pm320$  &
$2.24\pm 0.73$  & $11.45\pm 3.51$ & $26.4/19$ \\
\SiXIII ($fir$)  & 6.648   & $2.45\pm0.28$  & $-133\pm479$  & $1017\pm256$  &
$2.57\pm0.89$ & $11.03\pm3.09$ & $44.5/40$ \\
\SiXIV  & 6.180   & $1.11\pm 0.11$  & $ 50\pm255$   & $956\pm285$  &
$2.64\pm0.80$  & $18.85\pm 6.26$  & $15.9/14$ \\
\SXV ($fir$)  & 5.039   & $1.86\pm0.41$  & $ -832\pm368$   & $533\pm348$  &
$2.15\pm1.04$  & $16.18\pm 4.96$ & $21.0/24$ \\
\SXVI  & 4.727   & $0.52\pm0.12$  & $-35\pm635$   & $< 2050 $  &
$2.06\pm0.94$  & $29.32\pm 10.49$ & $6.2/10$ \\
\ArXVII ($fir$)  & 3.949   & $0.56\pm0.20$  & $-736\pm258$   & $< 650$  &
$3.74\pm2.06$  & $23.74\pm7.89$ & $21.1/22$ \\
\tableline
\end{tabular}
\end{center}
\tablenotetext{a}{\TL\ range limits corresponds to values in $T$ for 
which line emission is within 75\% of its most probable value}
\end{table}

\clearpage

\begin{table}[b]
\caption{He-like ion $fir$ Line Properties for \cygate 
\label{tab:fir}}
\vspace{-0.1cm}
\begin{tabular}{cccccccc}
\tableline
\tableline
Ion & $r$ flux & $i$ flux & $f$ flux & G-ratio & $f/i$ ratio & $T_{fir}$&
R$_{fir}$\\
 & & &  & & & (MK)  & \Rstar \\
\tableline
\multicolumn{8}{l}{\textit{\underline{Medium Energy Grating}} } \\    
\MgXI     & $0.45\pm0.04$  & $0.15\pm0.02$  & $0.05\pm0.01$   &
$0.48\pm0.07$  & $ 0.35\pm0.08$  &   $> 11$     & $3.13\pm0.39$  \\
\SiXIII   & $1.17\pm0.07$   & $0.46\pm0.04$  & $0.57\pm 0.05$   &
$0.87\pm0.10$  & $1.20\pm0.17$   &  $< 5 $      & $1.96\pm 0.24$\\
\SXV      & $0.60\pm0.10$  & $0.41\pm0.09$  & $0.35\pm0.08$    &
$1.26\pm0.44$  & $0.85\pm0.33$  &   $< 5$     & $ < 1.11$  \\
\ArXVII   & $0.12\pm0.05$  & $0.18\pm0.06$  & $< 0.03$   &
$1.57\pm1.04$  & $< 0.16$     & $< 25$   & $\sim 1.00$  \\
\multicolumn{8}{l}{\textit{\underline{High Energy Grating}} } \\      
\MgXI     & $0.52\pm0.07$  & $0.22 \pm0.05$ & $0.17\pm0.04$   &
$0.76\pm0.16$  & $0.80\pm0.21$  & $5.04\pm2.71$   & $5.13\pm0.93$ \\
\SiXIII   & $1.04\pm0.11$  & $0.61\pm0.08$  & $0.80\pm0.09$   &
$1.34\pm0.23$  & $1.25\pm0.24$  & $<3$  & $2.06\pm0.37$  \\
\SXV      & $0.58\pm0.13$  & $0.67\pm0.14$  & $0.61\pm0.14$   &
$2.20\pm0.87$   & $0.92\pm0.35$  &   $<4$    &  $<1.18$   \\
\ArXVII   & $0.28\pm0.09$  & $0.20 \pm0.07$  & $0.08\pm0.05$   &
$1.00\pm0.68$  & $ 0.42\pm0.37$      & $<100$   & $\sim 1.00$  \\
\tableline
\end{tabular}
\tablenotetext{a}{The $f$, $i$, $r$ line fluxes are in units $10^{-13}$ erg
cm$^{-2}$ s$^{-1}$}
\end{table}

\clearpage

\begin{table}[t]
\caption{H-like to He-like Line Ratio Temperatures for 
Cyg OB2 Nos. 8a and 9 \label{tab:MEG_HHe}}
\vspace{-0.1cm}
\begin{tabular}{ccccc}
\tableline
\tableline
Ions     & HETGS/MEG &  & HETGS/HEG &  \\
\tableline
     & Line Ratio & $T_{H/He}$ (MK)  & Line Ratio &  $T_{H/He}$ (MK)  \\
      \tableline
       \multicolumn{5}{l}{\textit{\underline{Cyg No. 8a}} } \\    
\MgXII / \MgXI(r) & $1.23\pm0.23$  &  $8.46\pm0.52$   &
$1.40\pm0.36$  & $8.78\pm0.77$ \\
\SiXIV / \SiXIII(r) & $1.01\pm0.13$  & $12.96\pm0.58$  & $1.17\pm0.23$  &
$13.61\pm0.99$ \\
\SXVI / \SXV(r)    & $0.60\pm0.24$  & $16.50\pm2.21$  & $0.91\pm0.42$  &
$18.90\pm3.32$ \\
\multicolumn{5}{l}{\textit{\underline{Cyg No. 9}} } \\      
\MgXII / \MgXI(r) & $0.94\pm0.39$  &  $7.70\pm1.02$   & ---  & --- \\
\SiXIV / \SiXIII(r) & $1.18\pm0.42$  & $13.60\pm1.74$  & ---  & --- \\
\tableline
\end{tabular}
\end{table}

\clearpage

\begin{figure}[!ht]
\rotatebox{0}{ 
\resizebox{15.2cm}{!}{\includegraphics{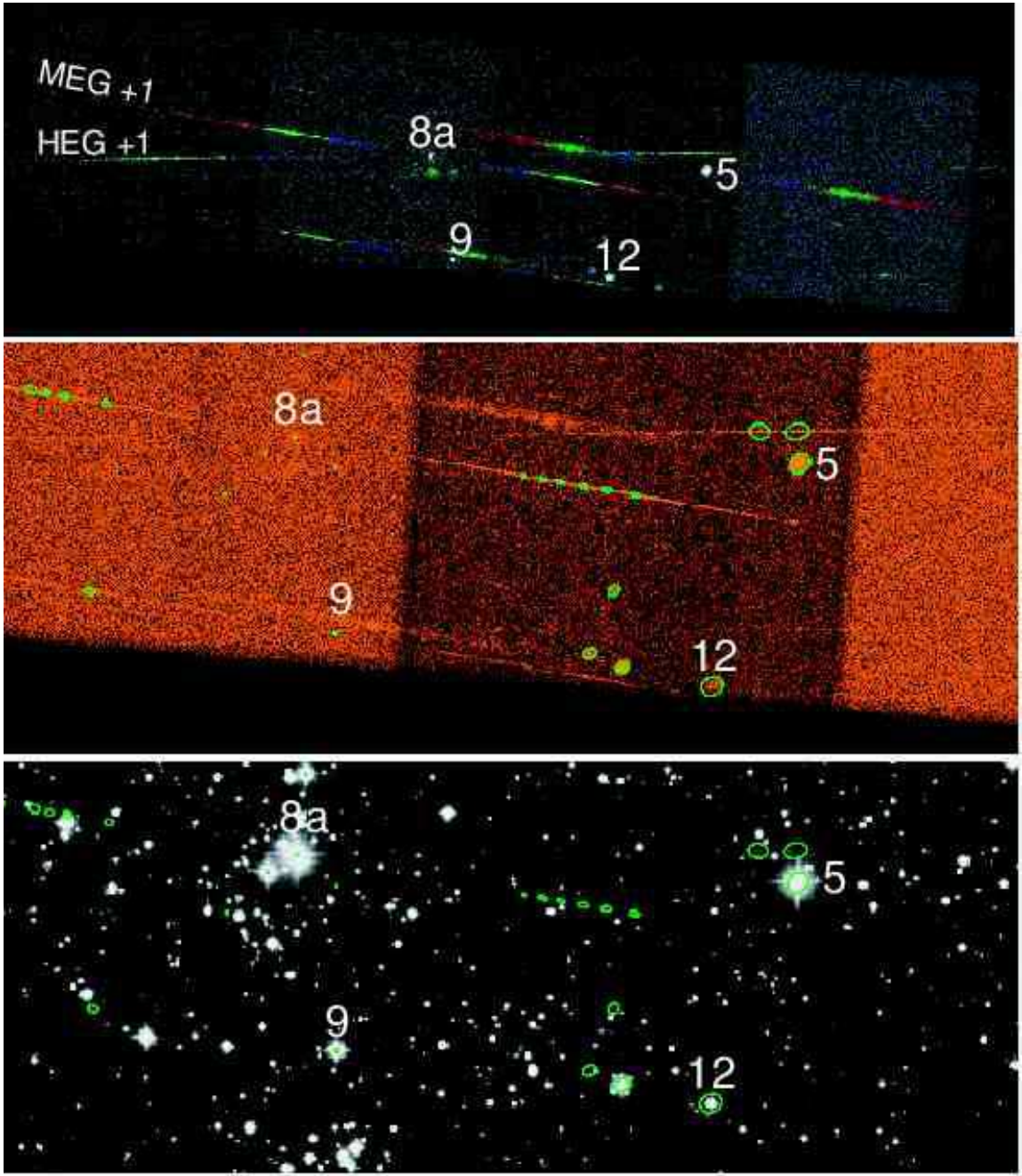}}}
\caption{This shows a ``true-color'' X-ray image showing the \Chandra\
spectra of the central region of Cyg OB2.  For all images, North is up.  Photons are color-coded
according to the following scheme: Red $< 1.5$ keV, Green = $1.5 -2.5$ keV,
and Blue = $2.5-8$ keV.   The middle panel shows the ACIS-I focal plane image
with the sources (shown by green circles) found using the CIAO tool TGDETECT.  The bottom
panel shows the same regions on a Digital Sky Survey of the same region.
Although some of these detected sources correspond to stars in the Cyg OB2
Association, several circled regions shown in the middle and bottom panels are not actual X-ray
sources.  For example, the straight rows of circled regions along the MEG spectrum on either side
of Cyg OB2 No. 8a and a few other circled regions (as determined by the automatic source
detection routine) are simply marking the locations of bright emission lines in the dispersed X-ray
spectra of the four brightest stars.
\label{Fig1}}
\end{figure}

\clearpage

\begin{figure}[!ht]
\vspace{3cm}
\rotatebox{0}{ 
\resizebox{12cm}{!}{\includegraphics{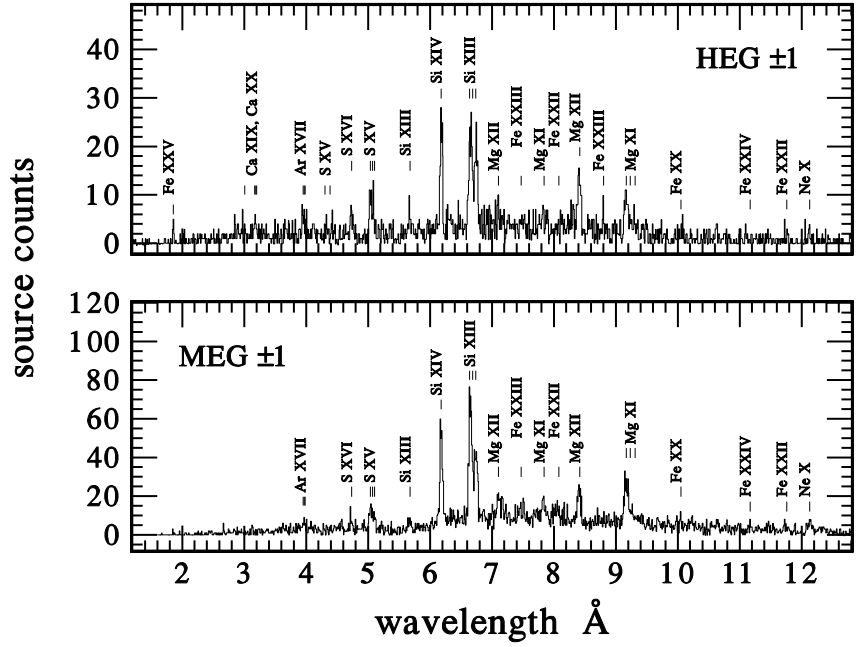}}}
\caption{Observed HEG and MEG $\pm1^{st}$ order count spectra for 
Cyg OB2 No. 8a.  The most likely line identifications are indicated.  
The bin size is 0.01 \AA.   
\label{HEGMEG1} }
\end{figure}

\clearpage

\begin{figure}[!ht]
\vspace{3cm}
\rotatebox{0}{ 
\resizebox{12cm}{!}{\includegraphics{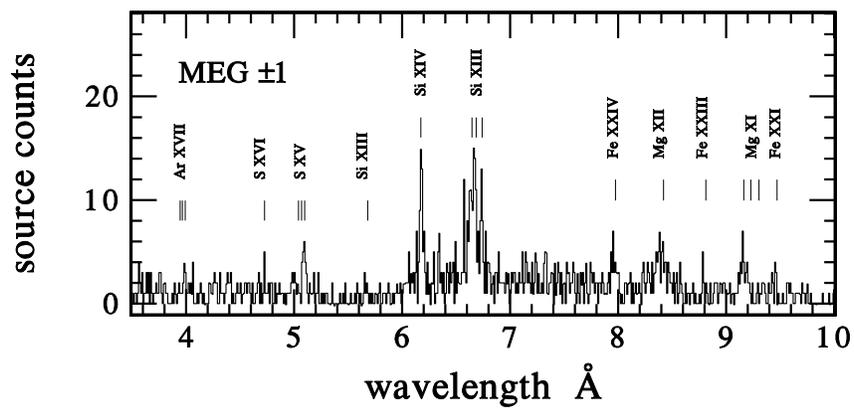}}}
\caption{Observed MEG $\pm1^{st}$ order count spectrum for 
Cyg OB2 No. 9.  The most likely line identifications are indicated.  
The bin size is 0.01 \AA. 
\label{MEGCYG9}}
\vspace{-0.2cm}
\end{figure}

\clearpage

\begin{figure}[!ht]
\rotatebox{-90}{ 
\resizebox{16cm}{!}{\includegraphics{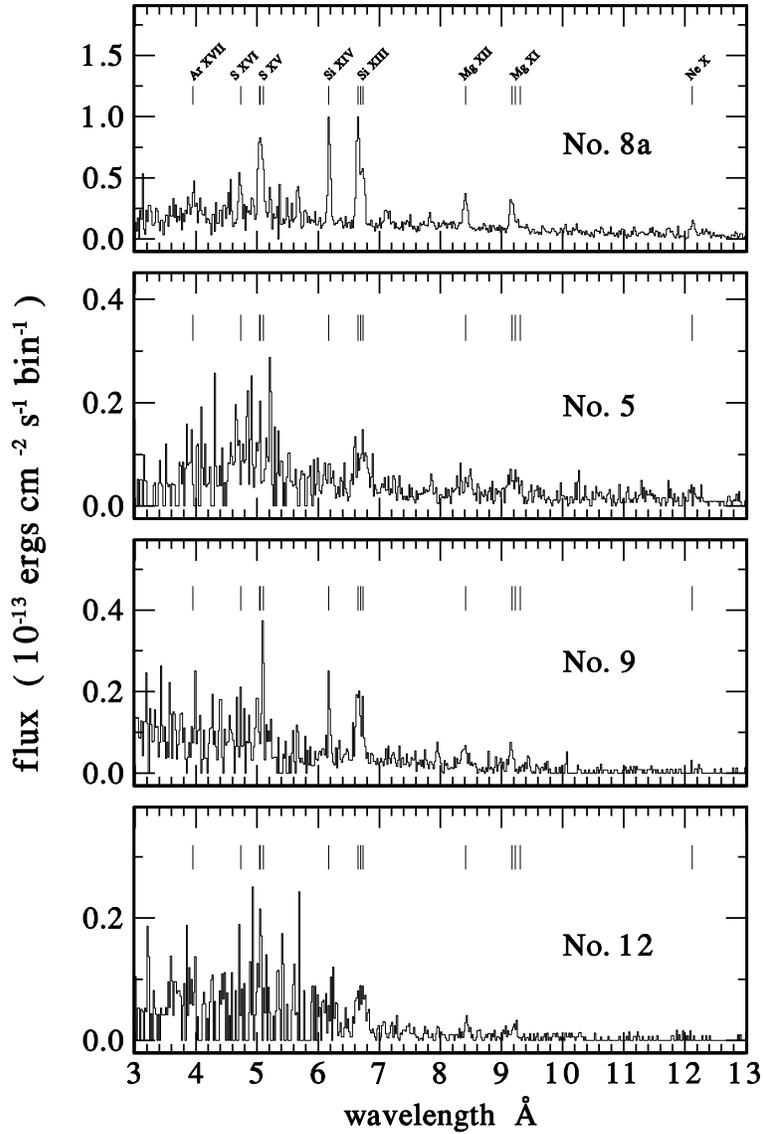}}}
\caption{Comparison of the combined MEG $\pm1^{st}$ 
order flux spectra for all four stars (Cyg OB2 Nos. 8a, 5, 9, \& 12).  The
rest wavelengths of the H-like and He-like lines are labeled.  The off-axis
degradation in the flux spectra of Cyg OB2 Nos. 5, 9, \& 12 is clearly
visible below 6 \AA.  The bin size is 0.02 \AA.
\label{Fig4} } 
\vspace{-0.1cm}
\end{figure}

\clearpage

\begin{figure}[!ht]
\rotatebox{-90}{ 
\resizebox{16cm}{!}{\includegraphics{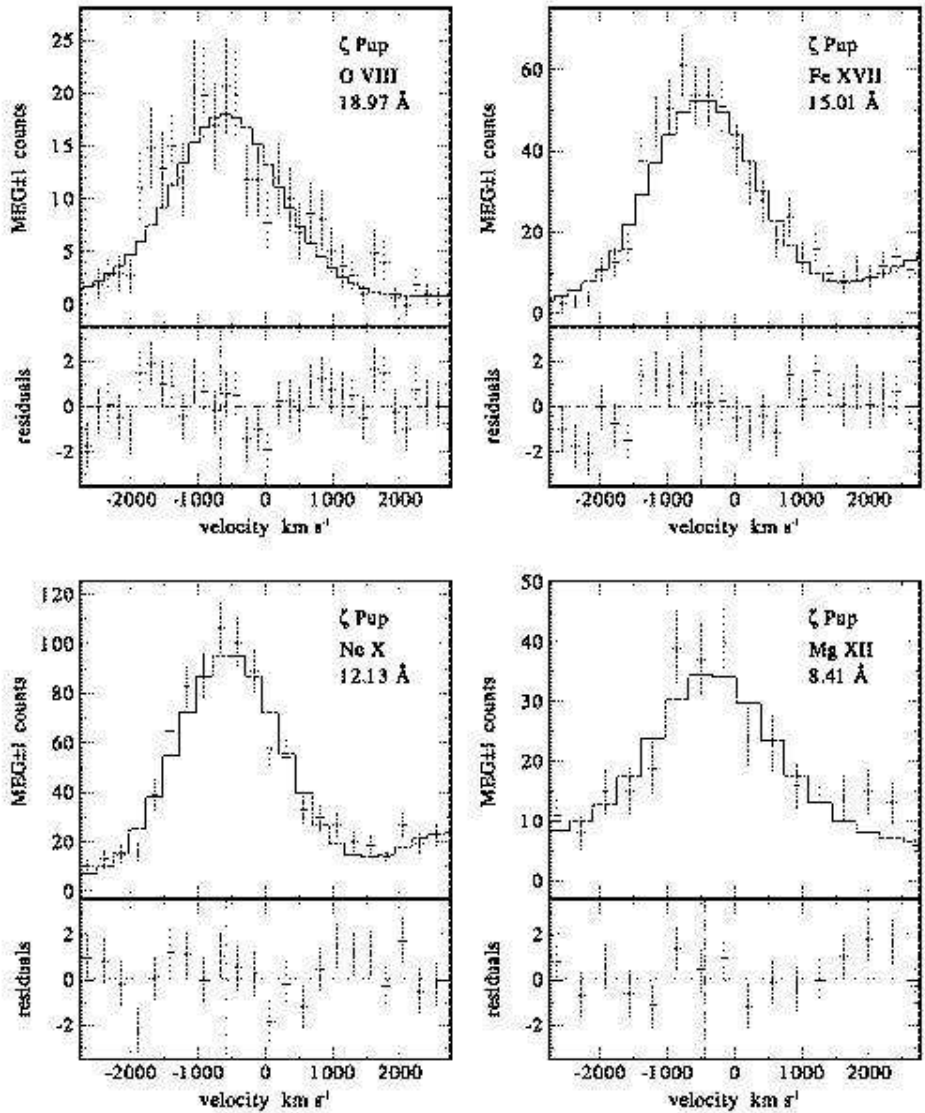}}}
\caption{Shows comparisons of the four strongest Zeta Pup H-like best-fit
line model spectra (solid-line) with their corresponding observed $MEG\pm1$
count spectra and their associated residuals.  The horizontal axis is
expressed as the velocity shift relative to the rest wavelength of the line.
The vertical dashed-lines in the residual plots represent the best-fit line
centroid shift velocities ($V_S$).  In order to obtain reliable flux
measurements of the line and continuum emissions for \NeX, the input model 
spectrum included the weaker  \FeXVII ($\sim 12.26$ \AA) located at
$\sim 3200$ \kms. The bin size is 0.01 \AA.
\label{ZPUPHLINES} }
\end{figure}

\clearpage

\begin{figure}[!ht]
\rotatebox{-90}{ 
\resizebox{16cm}{!}{\includegraphics{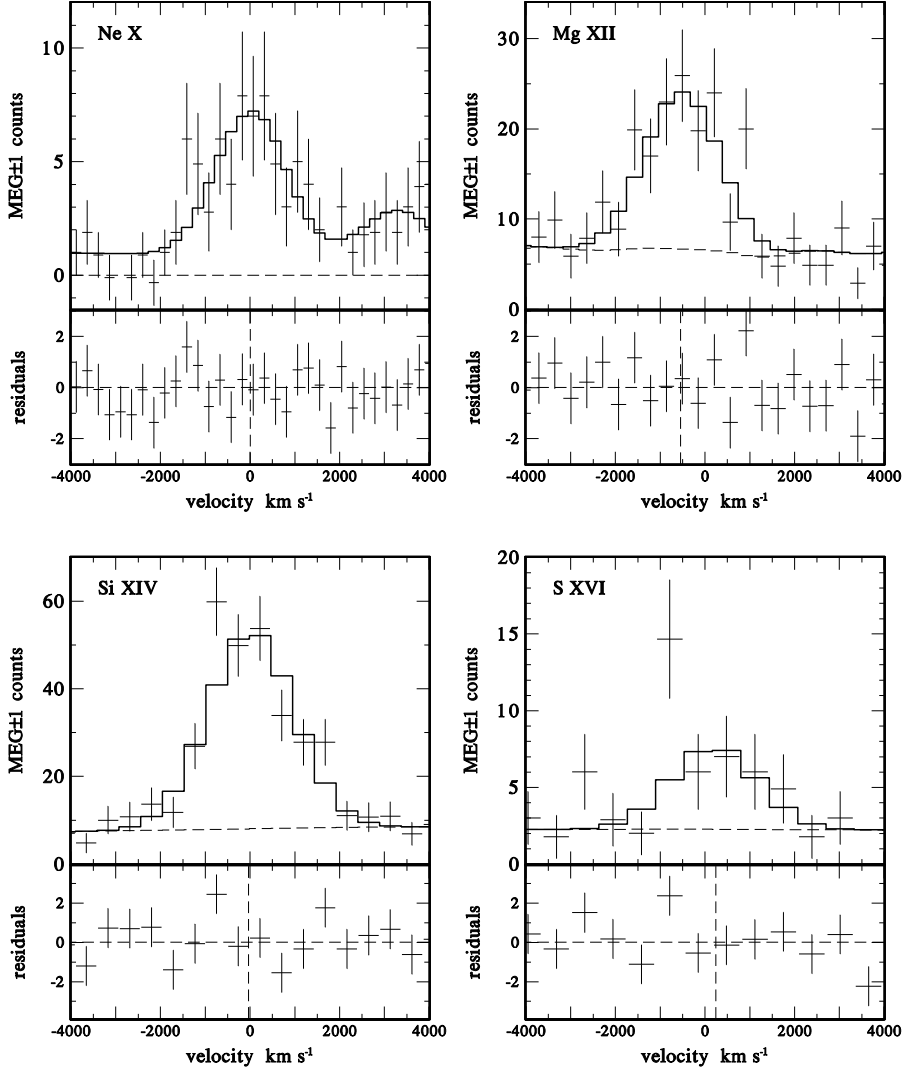}}}
\caption{Shows comparisons of the Cyg OB2 No. 8a H-like best-fit line model
spectra (solid-line) with their corresponding observed $MEG\pm1$ count
spectra and their associated residuals.  The horizontal axis is expressed as
the velocity shift relative to the rest wavelength of the line. The
horizontal dashed-line in the best-fit plot represents the predicted model
counts for the continuum. The vertical dashed-lines in the residual plots
represent the best-fit line centroid shift velocities ($V_S$).  In order to
obtain reliable flux measurements of the line and continuum emissions for
\NeX, the input model spectrum included the weaker  \FeXVII ($\sim 12.26$ \AA)
located at $\sim 3200$ \kms. The bin size is 0.01 \AA.
\label{MEGHLINES} }
\end{figure}

\clearpage

\begin{figure}[!ht]
\rotatebox{-90}{ 
\resizebox{16cm}{!}{\includegraphics{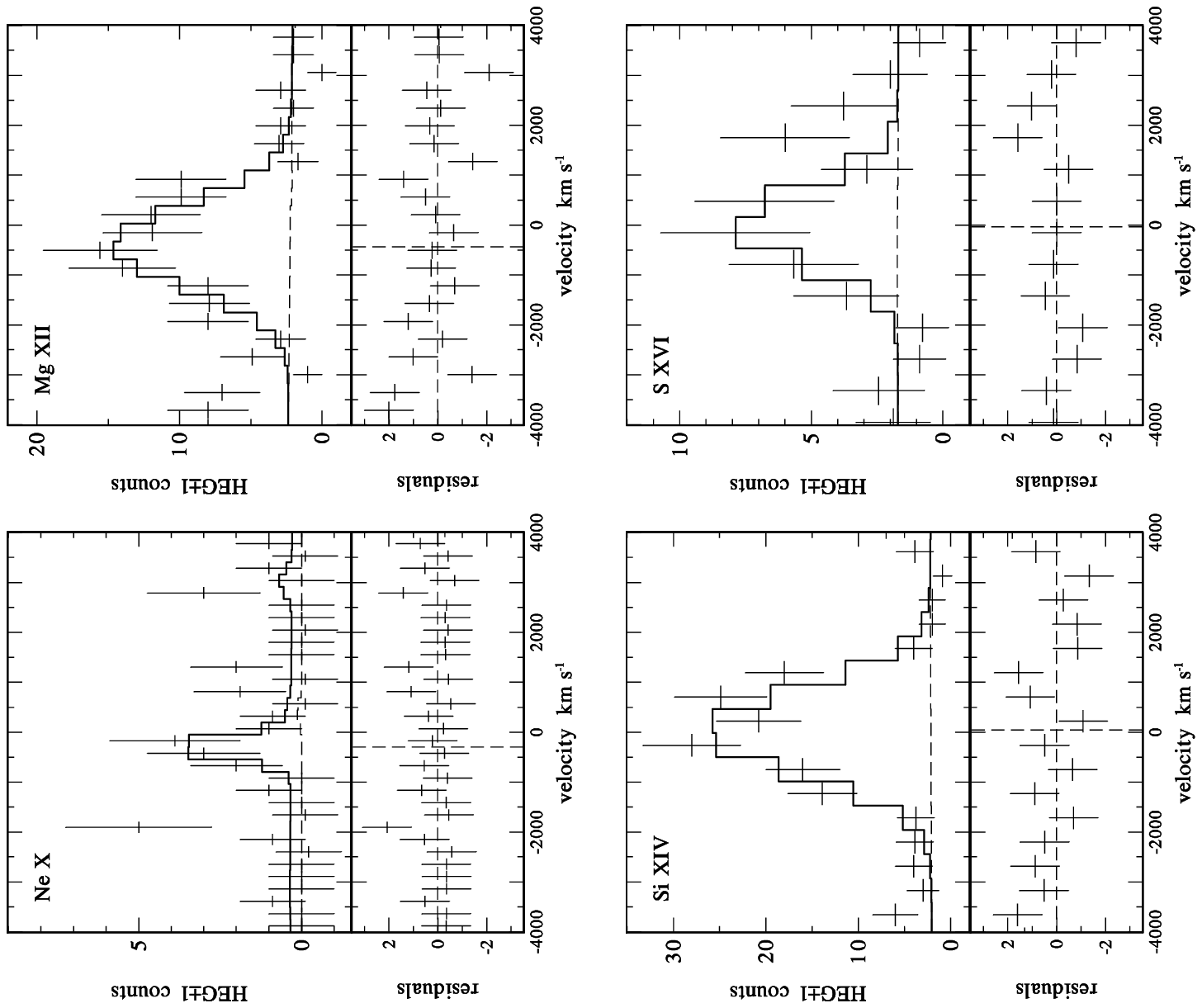}}}
\caption{Same as Figure 6 except for comparisons with the $HEG\pm1$
count spectra.
\label{HEGHLINES} }
\end{figure}

\clearpage

\begin{figure}[!ht]
\rotatebox{-90}{ 
\resizebox{16cm}{!}{\includegraphics{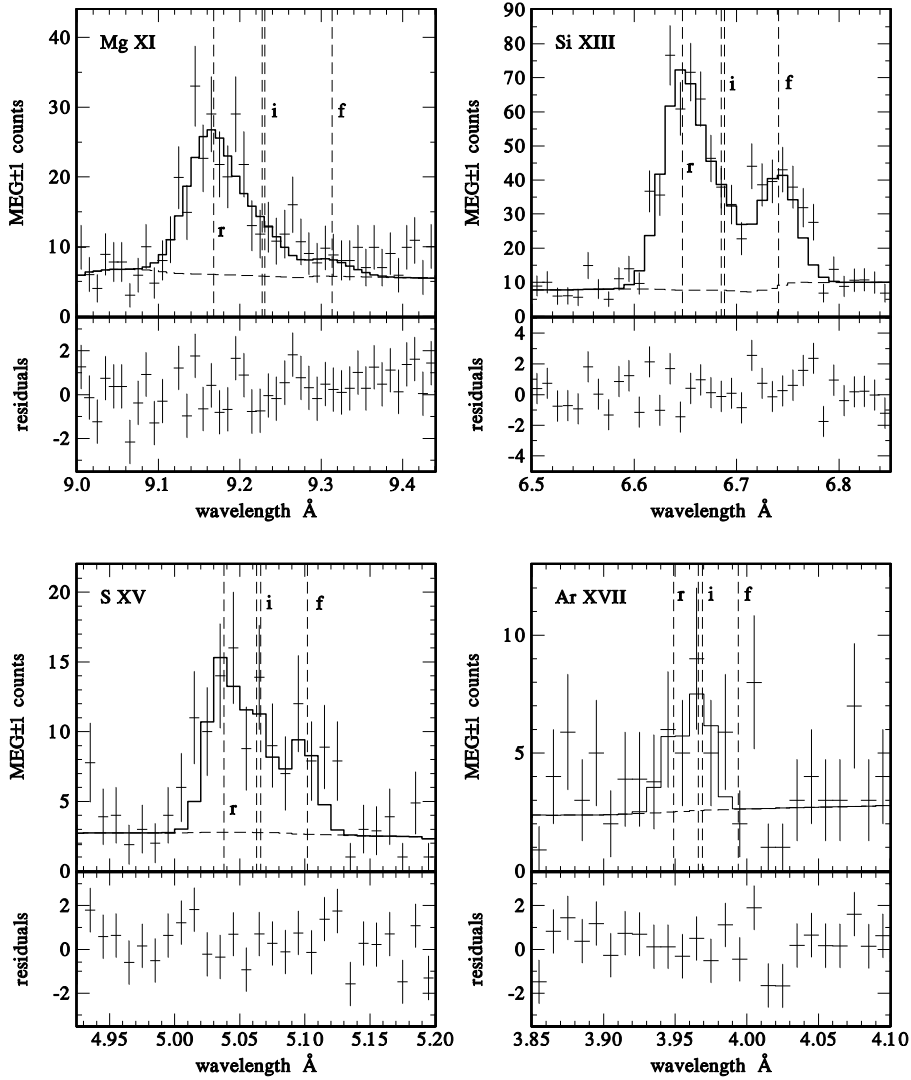}}}
\caption{Shows comparisons of the Cyg OB2 No. 8a He-like best-fit line
model spectra (solid- line) with their corresponding observed $MEG\pm1$
count spectra and their associated residuals. The vertical dashed-lines
indicated the rest wavelengths of the \fir\ lines. Note, although the $i$
line is actually a doublet, each input model spectrum assumes only one line.
The bin size is 0.01 \AA.
\label{MEGFIRLINES} }
\end{figure}

\clearpage

\begin{figure}[!ht]
\rotatebox{-90}{ 
\resizebox{16cm}{!}{\includegraphics{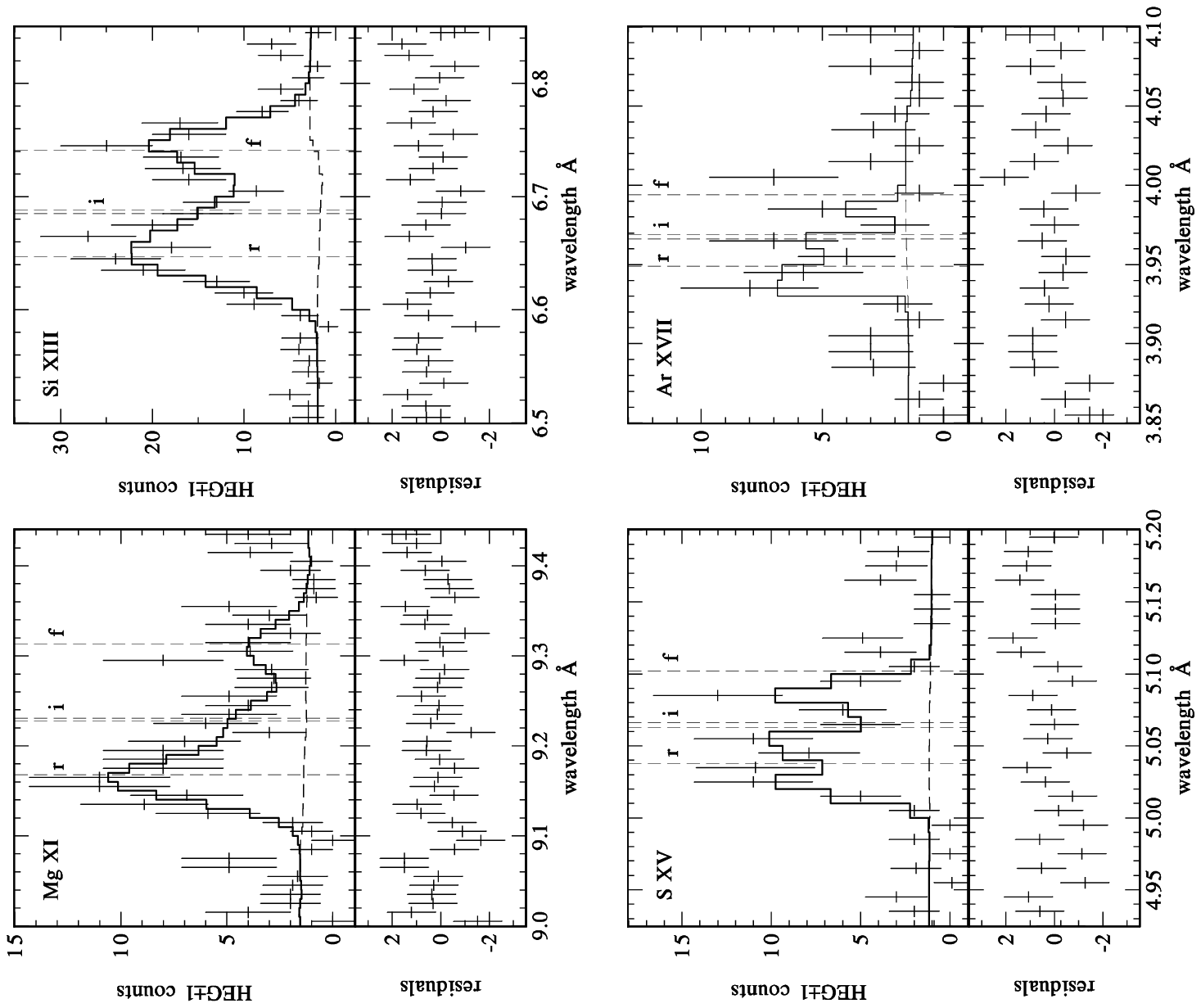}}}
\caption{Same as Figure 8 except for comparisons with the $HEG\pm1$
count spectra.
\label{HEGFIRLINES} }
\end{figure}

\clearpage

\begin{figure}[!ht]
\rotatebox{-90}{ 
\resizebox{16cm}{!}{\includegraphics{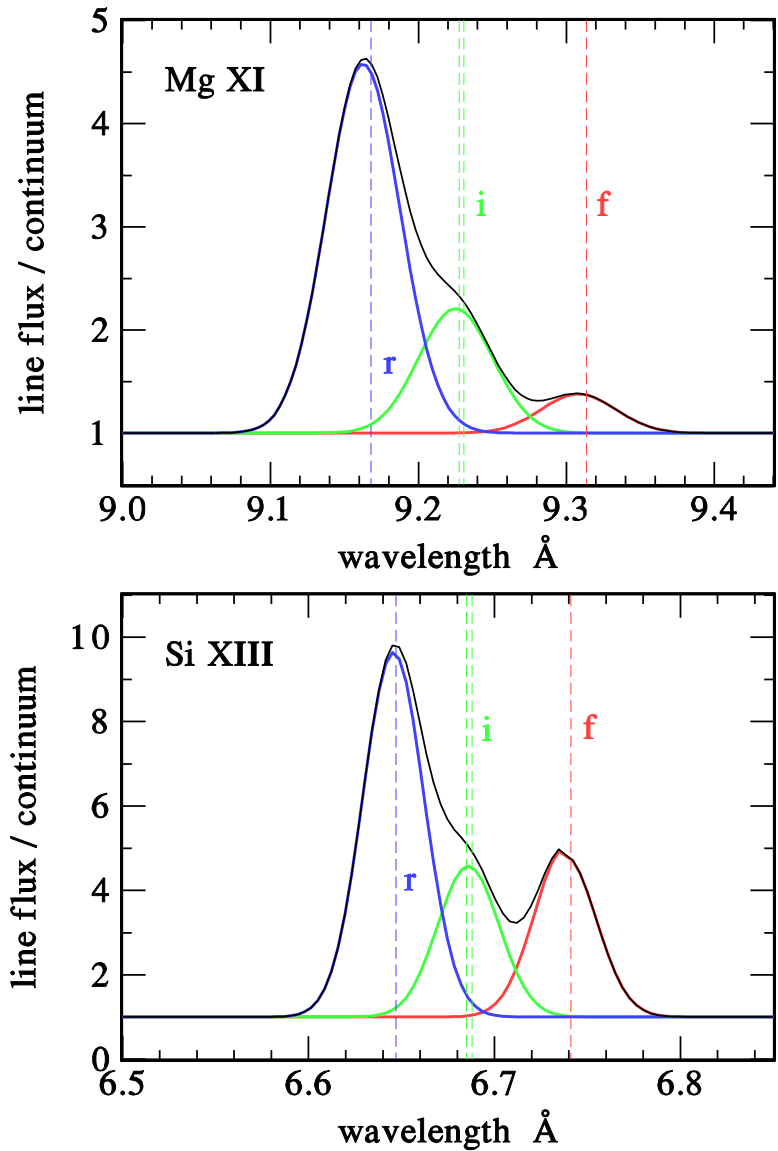}}}
\caption{Shows the best-fit input model spectra for two He-like \fir\ line
triads (\MgXI \& \SiXIII) prior to folding through the $MEG\pm1$
instrumental response functions (ARF \& RMF).  The input spectra are
normalized by their respective continuum.  The contributions of each line to
the total (black line) line emission is indicated: r-line (blue), i-line
(green), and f-line (red).  This demonstrates that the high energy
resolution capabilities of the HETGS provides a clear distinction of each
line's contribution to the overall emission, and allows us to extract
individual \fir\ line characteristics.
\label{MODELHELINES} }
\end{figure}

\clearpage

\begin{figure}[!ht]
\vspace{3cm} 
\rotatebox{0}{
\resizebox{16cm}{!}{\includegraphics{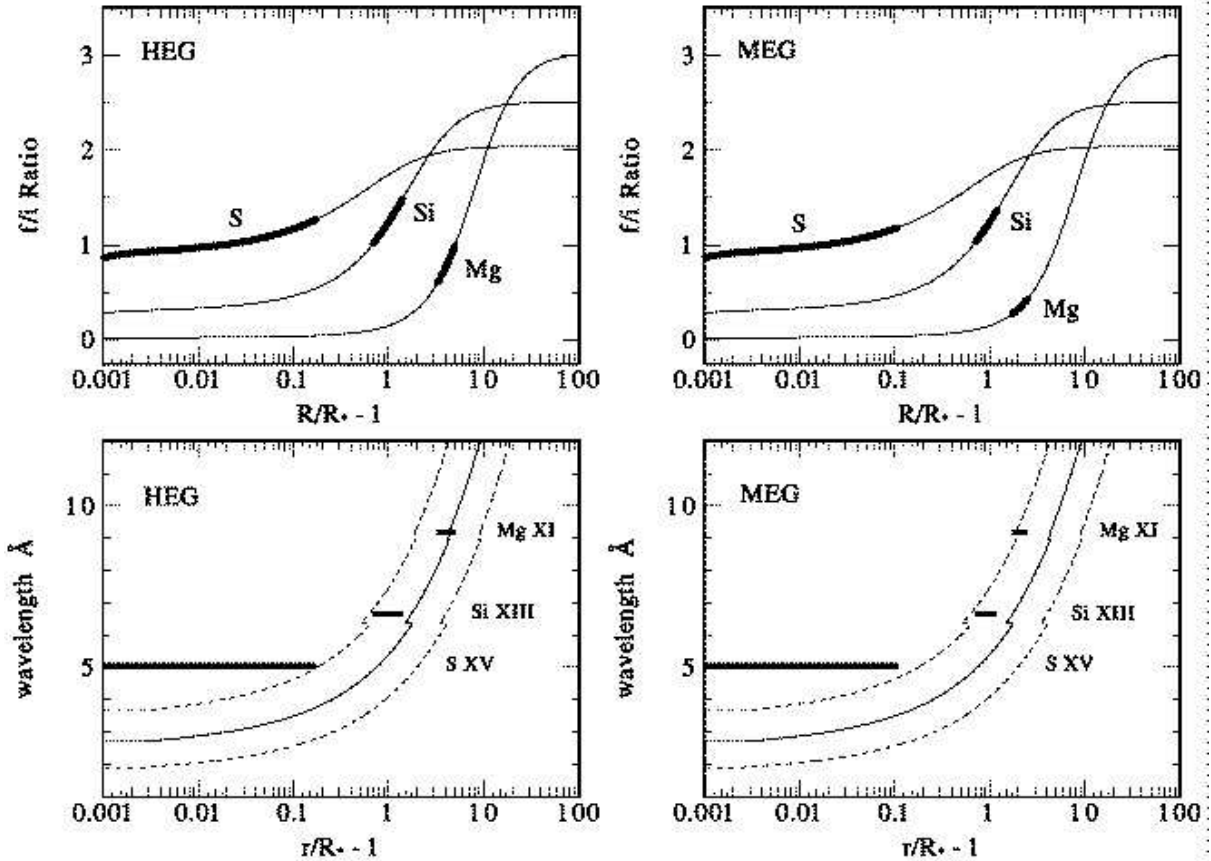}}}
\caption{ Top - The Cyg OB2 No. 8a HEG and MEG He-like $f/i$ ratio dependence on
radius.  The observed range of the $f/i$ ratios and their associated radial ranges are indicated by
the broad darkened line sectors.  Bottom - The
corresponding X-ray continuum optical depth unity (solid line) wavelength
dependence on radius (assuming the mass loss rate given in Table 2) and the
observed range in radii associated with each He-like $f/i$ ratio.  In all
plots the \ArXVII\ radial range is not shown since it is below 1.001. The
dashed lines correspond to a factor of 2 increase (larger radii) and a
factor of 2 decrease (smaller radii) in mass loss rate.}
\label{FIRTAU1}
\end{figure}

\clearpage

\begin{figure} [!ht]
\rotatebox{-90}{ 
\resizebox{16cm}{!}{\includegraphics{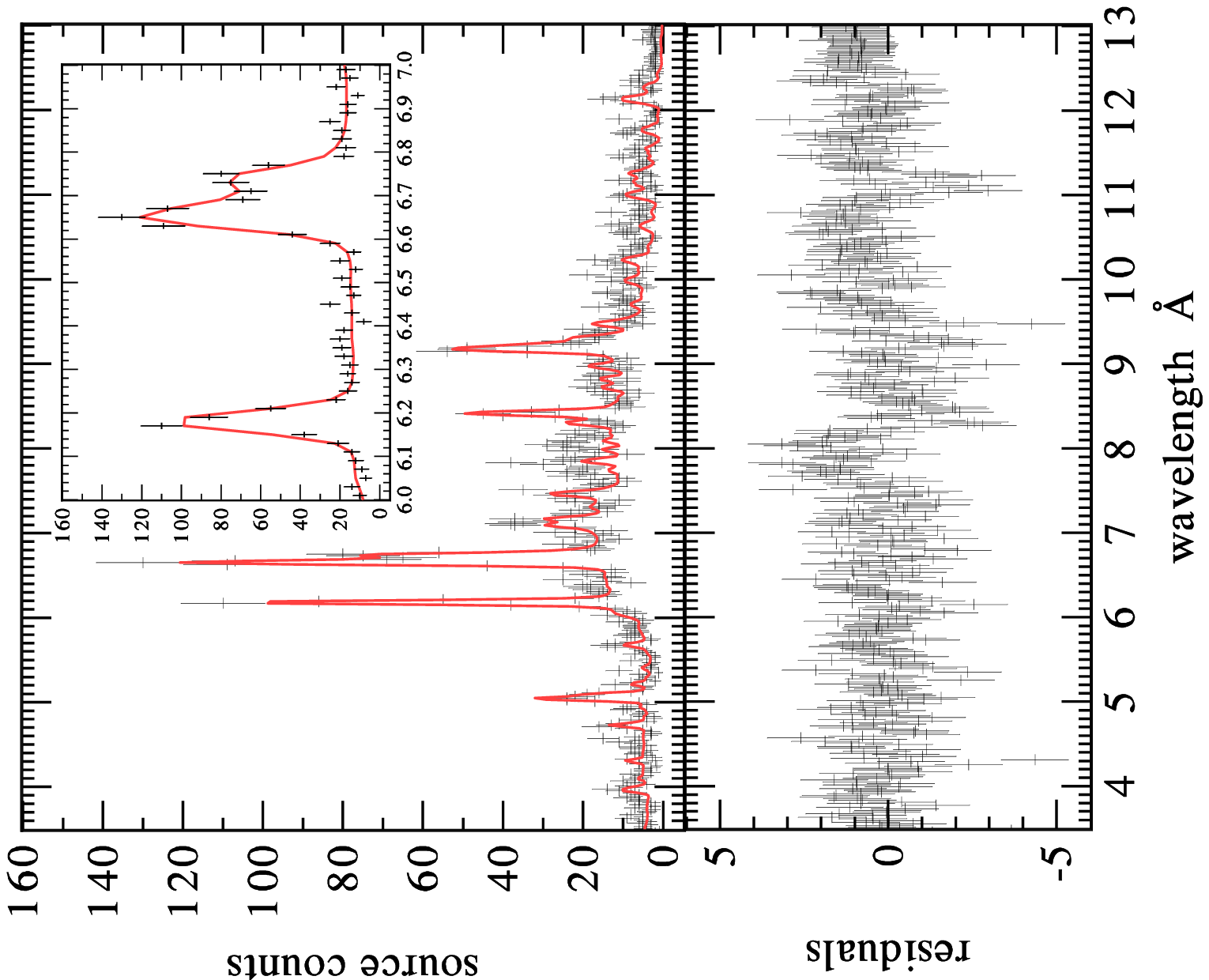}}}
\caption{Comparison of the best-fit model spectrum (red line) to the
MEG $\pm 1^{st}$ order spectrum for Cyg OB2 No. 8a.  The spectrum only
covers the wavelength region of the strongest lines. The corresponding
residuals of the fit are shown in the bottom panel.  The inset highlights
the wavelength region of the two strongest lines, \SiXIV\ and \SiXIII. The
bin size is 0.02 \AA.
\label{BESTFIT} }
\vspace{-0.6cm}
\end{figure}


\vspace{1cm}

\begin{references}
\reference{}Abbott, D. C., Bieging, J. H., \& Churchwell, E. 1981, 
         \apj, 250, 645
\reference{}Abbott, D. C., Telesco, C. M., \& Wolff, S. C. 1984, 
         \apj, 279, 225
\reference{}Bergh\"{o}fer, T. W., \& Schmitt, J. H. M. M. 1994, 
         A\&A., 290, 435 
\reference{}Bergh\"{o}fer, T. W., Schmitt, J. H. M. M., Danner, R., 
        \& Cassinelli, J. P.  1997,A\&A., 322, 167 
\reference{}Bieging, J. H., Abbott, D. C., \& Churchwell, E. B. 1989, 
        \apj, 340, 518
\reference{}Blumenthal, G.\ R., Drake, G.\ W.\ F., \& Tucker, W.\ H.\ 1972,
     \apj, 172,  205 
\reference{}Bonnell, I. A., \& Bate, M. R. 2002, MNRAS, 336, 659
\reference{}Bonnell, I. A., Vine, S. G., \& Bate, M. R. 2004, MNRAS, 349, 735
\reference{}Cassinelli, J. P., \& Olson, G. L. 1979, \apj, 229, 403
\reference{}Cassinelli, J. P., Waldron, W. L., Sanders, W., Harnden, F., 
      Rosner, R., \&  Vaiana, G.  1981, \apj\ 250,677 
\reference{}Cassinelli, J. P., Miller, N. A., Waldron, W. L., 
            MacFarlane, J. J., \&  Cohen, D. H. 2001, ApJ 554, L55
\reference{}Cassinelli, J. P., \& Swank, J. H. 1983, \apj, 271, 681
\reference{}Cassinelli, J. P., Waldron, W.  L., \& Miller, N. A. 2003, NASA MSFC Symp., 
      4 Years of Chandra Observations: A Tribute to Riccardo Giacconi, 
      Sept. 16-18
\reference{}Chandra X-Ray Center (CXC) 2002, Chandra Proposers' 
      Observatory Guide, Rev. 5.0, 2002, TD 403.00.005 (Cambridge, MA: CXC)
\reference{}Chlebowski, T. 1989, \apj, 342, 1091
\reference{}Cooper, G. 1994, Ph.D. Thesis, University of Delaware
\reference{}Cohen, D., Cassinelli, J. P., \& MacFarlane J. J. 1997, \apj, 487, 687
\reference{}Corcoran, M.\ F.\ et  al.\ 1993, \apj, 412, 792
\reference{}Feldmeier, A.\ 1995, \aap, 299, 523 
\reference{}Feldmeier, A., Shlosman, I., \& Hamann, W. -R. 2002, 
      \apj, 566, 392
\reference{}Feldmeier, A., Oskinova, L., \& Hamann, W. -R. 2003, A\&A, 403, 217 
\reference{}Groenewegen, M. A. T., Lamers, H. J. G. L. M., \& Pauldrach, A. W. A. 1989, \aap,  
     221, 78
\reference{}Hanson, M.\ M.\ 2003, \apj, 597, 957 
\reference{}Harnden, F.\ R.\etal\ 1979, \apjl, 234, L51 
\reference{}Herrero, A., Puls, J., \& Najarro, F. 2002, A\&A, 296, 949
\reference{}Howk, J. C., Cassinelli, J. P., Bjorkman, J. E., \& Lamers, H. J. G. L. M. 2000, \apj,   
     534, 348
\reference{}Ignace, R., \& Gayley, K. 2002, \apj, 568, 594
\reference{}Ignace, R., Oskinova, L. M., \& Brown, J. C. 2003, A\&A, 408, 353
\reference{}Kahn, S.\ M., Leutenegger, M.\ A., Cottam, J., Rauw, G., 
     Vreux, J.-M., den Boggende, A.\ J.\ F.,  Mewe, R., \& 
     G{\"u}del, M.\ 2001, \aap, 365, L312 
\reference{}Kitamoto, S. \& Mukai, K. 1996, PASJ, 48, 813
\reference{}Kn\"{o}dlseder, J. 2000, A\&A, 360, 539
\reference{}Kramer, R. H., Cohen, D. H., \& Owocki, S. P. 2003, \apj, 592, 532.
\reference{}Lamers, H. J. G. L. M. \& Cassinelli, J. P. 1999, ``Introduction to Stellar Winds``,
       Univ. Cambridge Press
\reference{}Lucy, L.\ B.\ \& White, R.\ L.\ 1980, \apj, 241, 300 
\reference{}MacFarlane, J.\ J., Cassinelli, J.\ P., Welsh, B.\ Y., 
       Vedder, P.\ W., Vallerga, J.\ V., \&  Waldron, W.\ L.\ 1991, 
       \apj, 380, 564 
\reference{}MacFarlane, J.\ J., Waldron, W.\ L., Corcoran, M.\ F., 
      Wolff, M.\ J., Wang, P., \& Cassinelli,  J.\ P.\ 1993, \apj, 419, 813 
\reference{}Mewe, R., Gronenschild, E. H. B. M., \& 
      van den Oord, G. H. J. 1985, A\&AS, 62,  197
\reference{} Miller, N. A., Cassinelli, J. P., Waldron, W. L., 
      MacFarlane, J. J., \& Cohen, D. H. 2002, \apj, 577, 951
\reference{} Miller, N. A., 2002, Ph.D. Thesis, University of Wisconsin.
\reference{} Moeckel, N., Cho, J., \& Cassinelli, J. P. 2002, BAAS, 200, 7418
\reference{} Porquet, D., \& Dubau, J. 2000, \aap, 143, 495 
\reference{}Owocki, S.\ P., Castor, J.\ I., \& Rybicki, G.\ B.\ 1988, 
    \apj, 335, 914 
\reference{}Owocki, S.\ P., \& Cohen, D.\ H. 2001 \apj, 559, 1108
\reference{}Raymond, J. C. 1988, Hot Thin Plasmas in Astrophysics, ed., R. Pallavicini,
(Dordrecht: Kluwer), p. 3
\reference{}Raymond, J. C., \& Smith, B. W. 1979, \apjs, 35, 419
\reference{}Schaller, G., Shaerer, D., Meynet, G., \& Maeder, A. 1992, 
    A\&AS, 96, 269
\reference{}Schulz, N. S., Canizares, C. R., Huenemoerder, D., \& 
    Tibbets, K. 2003, \apj, 595, 365 
\reference{} Shull, J. M., \& Van Steenberg, M. E. 1985, \apj, 294, 599.
\reference{} Smith, R. K., \& Brickhouse, N. S. 2000, Rev. Mexicana 
    Astron. Astrofis. Ser. Conf., 9, 134
\reference{}Snow, T., \& Morton D. 1977, \apjs, 33, 269
\reference{} Underhill, A. B. 1980, \apj, 239, 414
\reference{}Waldron, W.  L. 1984, \apj, 282, 256 
\reference{}Waldron, W.  L., \& Cassinelli, J. P. 2001 \apj, 548, L45
\reference{}Waldron, W.  L., \& Cassinelli, J. P. 2002, ASP Conf. Ser., 
   The High Energy Universe at Sharp Focus: Chandra Science, 
   Vol. 262, eds., E. M. Schlegel \& S. Dil Vrtilek, (San Francisco), p. 69
\reference{}Waldron, W. L., Corcoran, M. F., Drake, S. A., \& 
      Smale, A. P. 1998,\apjs, 118, 217 
\end{references}
\end{document}